\documentclass[a4paper,10pt,times,twocolumn,preprint,5p]{elsarticle}

\usepackage{blindtext}
\usepackage{graphicx}
\usepackage[font=scriptsize,labelfont=bf,justification=justified]{caption}
\usepackage{hyperref}
\usepackage{amssymb}
\usepackage{todonotes}

\usepackage{color}
\usepackage{soul}
\newcommand{\hlc}[2][white]{{%
    \colorlet{foo}{#1}%
    \sethlcolor{foo}\hl{#2}}%
}

\usepackage[ruled,linesnumbered]{algorithm2e}

\usepackage{amsmath}
\usepackage{arydshln}
\usepackage{verbatim} 
\usepackage{subfig} 
\usepackage{svg}
\usepackage{amsfonts}
\usepackage{amsopn}
\usepackage{amsthm}
\usepackage{xifthen}
\usepackage{pgfkeys}
\usepackage{multirow}
\usepackage{ragged2e} 
\usepackage{xfrac}
\usepackage{pdfpages}
\usepackage{natbib}
\usepackage{cleveref}
\usepackage{float}
\usepackage[export]{adjustbox}
\usepackage{cleveref}
\usepackage{enumitem}
\usepackage{tabularx}

\usepackage{booktabs} 
\usepackage{ltablex}
\usepackage{listings}
\usepackage{url}
\usepackage{adjustbox}

\begin{document}
\begin{frontmatter}
%%%%
%%%%
%%%%
\title{A Self-Integration Testbed for Decentralized Socio-technical Systems}
%%%%
%%%%
%%%%
\author[1]{\corref{correspondingAuthor}Farzam Fanitabasi}
%%%%
\cortext[correspondingAuthor]{Corresponding author. Stampfenbachstrasse 48, 8092 Zurich, Switzerland. Email: \href{mailto:farzamf@ethz.ch}{\textit{farzamf@ethz.ch}}}
%%%%
\author[1]{Edward Gaere}
%%%%
\author[2]{Evangelos Pournaras}
%%%%
\address[1]{Professorship of Computational Social Science, ETH Zurich, Zurich, Switzerland} 
\address[2]{School of Computing, University of Leeds, Leeds, UK}
%%%%
%%%%
%%%%
\begin{abstract}
The Internet of Things (IoT) comes along with new challenges for experimenting, testing, and operating decentralized socio-technical systems at large-scale.
\hlc{In such systems, autonomous agents interact locally with their users, and remotely with other agents to make intelligent collective choices. Via these interactions they self-regulate the consumption and production of distributed (common) resources, e.g., self-management of traffic flows and power demand in Smart Cities.}
While such complex systems are \hlc{often} deployed and operated using centralized computing infrastructures, the socio-technical nature of these decentralized systems requires new value-\hlc{sensitive} design paradigms;
empowering trust, transparency, and alignment with citizens' social values, such as privacy preservation, autonomy, and fairness among citizens' choices. 
Currently, instruments and tools to study such systems and guide the prototyping process from simulation, to live deployment, and ultimately to a robust operation of a high Technology Readiness Level (TRL) are missing, or not practical in this distributed socio-technical context. 
This paper bridges this gap by introducing a novel testbed architecture for decentralized socio-technical systems running on IoT. 
This new architecture \hlc{is designed for a} seamless reusability of 
(i) application-independent decentralized services by an IoT application, and 
(ii) different IoT applications by the same decentralized service. 
This dual self-integration promises IoT applications that are simpler to prototype, and can interoperate with decentralized services during runtime to self-integrate more complex functionality, e.g., data analytics, distributed artificial intelligence.
\hlc{Additionally, such integration provides stronger validation of IoT applications, and improves resource utilization, as computational resources are shared, thus cutting down deployment and operational costs.}
\hlc{Pressure and crash tests} during continuous operations of several weeks, with more than 80K \hlc{network joining and leaving of agents}, 2.4M parameter changes, and 100M communicated messages, confirm the robustness and practicality of the testbed architecture. 
This work promises new pathways for managing the prototyping and deployment complexity of decentralized socio-technical systems running on IoT, whose complexity has so far hindered the adoption of value-sensitive self-management approaches in Smart Cities.
\end{abstract}
%%%%
%%%%
%%%%
\end{frontmatter}
%%%%
%%%%
%%%%
\section{Introduction}
\label{S:Intro}
%%%%%
%%%%%
%%%%%
\hlc{The Internet of Things (IoT) radically transforms how complex socio-technical systems are designed, operated and managed. 
Smart Cities turn into organic ecosystems of ubiquitous sensors, autonomous vehicles, and personal pervasive devices that are massively interconnected and distributed}~\cite{arasteh2016iot,chernyshev2017internet,girau2016lysis}. 
New opportunities arise to control and manage socio-technical systems in real-time as the means to cope with uncertainties and continuous change~\cite{bellman2018self,bellman2019self,tomforde2020adapt}:
self-improving socio-technical operations by seamlessly self-integrating decentralized services that measure, learn, optimize, and adapt~\cite{pournaras2017self,pournaras2018decentralized,pournaras2017engineering}, i.e. load-balancing transport or power networks to prevent traffic congestion and blackouts.
However, the IoT complexity, heterogeneity, scale, infrastructural cost, and privacy concerns have so far limited \hlc{the} broader experimentation and research on designing such general-purpose services~\cite{chernyshev2017internet,gluhak2011survey}.
Nevertheless, \hlc{decentralized systems exhibit properties that can empower values by design in a socio-technical context:}
\hlc{(i) They can better preserve privacy by processing sensitive information locally and allowing informational self-determination}~\cite{zhang2005distributed,mahboubi2019privacy}.
\hlc{(ii) They are more transparent against algorithmic nudging and manipulation, as data are not centrally located and users preserve their autonomy}~\cite{preece2018asking,adadi2018peeking}.
\hlc{(ii) They can be designed to promote social welfare such as fairness}~\cite{huang2019embedding,fanitabasi2019appliance}.
Therefore, their adoption in socio-technical IoT applications of Smart Cities has a social and sustainability impact. \\

%In addition to the mentioned challenges, interconnected systems, such as IoT, are growing ever more complex~\cite{bellman2019self,}, due to the embedding of ICT functionality in more and more devices~\cite{bellman2018self}, increasingly dynamic operation environments, and internal or external disruptions~\cite{tomforde2020adapt}.
%This trend toward open, interconnected systems that (self-)integrate heterogeneous and autonomous devices and services at runtime, requires novel and improved modeling, analysis, simulation, and prototyping~\cite{bellman2019self,barnes2019chariot,tomforde2016organic}.
%Within such dynamic environments and during system runtime, devices can fail, users might join/leave, communication across devices and agents become disrupted, system goals, resources, and requirement can vary, and new services are required.
%Not all such changes can be foreseen during the initial design phase, and often it is infeasible for a centralized controller to know about all such changes in a timely manner~\cite{bellman2019self}.
%Thus, the challenge is to integrate and adapt to such changes in a rapidly and robustly.
%To address this, the ``self-improving system integration" (SISSY) initiative proposes \emph{self-integration}, defined as \emph{``an ongoing autonomous process for linking a potentially large set of heterogeneous computing systems, devices, and software applications; so as to meet system goals."}~\cite{bellman2019self}.

\hlc{Prototyping and testing decentralized socio-technical systems that continuously change and adapt is a challenge.  In particular, testing in real-world self-improving system integration (SISSY)}~\cite{bellman2018self,bellman2019self,tomforde2020adapt} \hlc{as the means to cope with complex system dynamics as well as user and network uncertainties remains to a high extent an ad hoc process.} This paper introduces a new IoT testbed architecture with a novel dual self-integration capability: 
(i) An IoT application integrates several application-independent and modular decentralized services to compose low-cost complex functionalities without changing the \hlc{application implementation.}
(ii) A decentralized service is integrated into several IoT applications without changing the \hlc{service implementation.}
This reusability is made possible by abstracting the software engineering complexity and interactions within two software agents under user's control: the \emph{application agent}, and \emph{service agent}. \\

Prototyping IoT applications and services to support, in particular, multiple self-integration scenarios, requires testing and refinements at multiple stages that start from simulations, move to live deployments, and ultimately to high Technical Readiness Level (TRL)\footnote{\href{https://www.nasa.gov/directorates/heo/scan/engineering/technology}{https://www.nasa.gov/directorates/heo/scan/engineering/technology} [Last accessed: May 2020]} operations. 
\hlc{Maintaining different implementations, or changing the code back and forth to validate new functionality is costly and complex}~\cite{fortino2017agent,girau2016lysis}. 
Experience shows that such flexibility for decentralized multi-agent systems is \hlc{extremely scarce}~\cite{arasteh2016iot,chernyshev2017internet}. \hlc{Existing toolkits cannot serve in practice the self-integration scenarios envisioned.}
This barrier is overcome by introducing a prototyping toolkit that extends and improves earlier work~\cite{galuba2009protopeer}: The \emph{\hlc{Livepeer}} toolkit. It provides support for IoT devices, i.e. software agent running on smart phones, a new efficient networking module, a new scalable logging infrastructure for system monitoring and analysis, as well as an improved design to limit earlier severe memory leaks and synchronization problems.  \\
%The \emph{Distributed and Intelligent Social Computing} (DISC) toolkit. 

The testbed architecture with the two software agents is experimentally evaluated with real-world data under long-lasting pressure and crash tests over several weeks. 
The complex operations of two decentralized socio-technical services are integrated in \hlc{Livepeer} as a proof of concept: 
(i) I-EPOS (\emph{Iterative Economic Planning and Optimized Sections})~\cite{pournaras2018decentralized}, and 
(ii) DIAS (\emph{Dynamic Intelligent Aggregation Service})~\cite{pournaras2017engineering}. 
I-EPOS performs decentralized combinatorial optimization using learning agents with structured interactions. 
In contrast, DIAS performs real-time collective measurements over a dynamic unstructured network of agents\hlc{, where agents can arbitrary join, leave, or fail, while their input data continuously change.} 
Both services empower highly sophisticated IoT application scenarios, such as traffic flow optimization, power peak-shaving, load-balancing of bike sharing stations, participatory crowd-sensing of mobility, and traffic~\cite{gerostathopoulos2019trapped,fanitabasi2019appliance,pournaras2018decentralized,pournaras2017engineering}. 
Results confirm the self-integration capability, and the performance benchmarks validate the robustness of the testbed architecture in scenarios of continuous change and adaptation, with more than 80,000 agents \hlc{joining and leaving the network}, 2.4 million parameter changes, and 100 million communicated messages.
The findings of this paper provide new insights to communities, government bodies, system operators and utilities on how to manage, operate, and regulate complex socio-technical IoT infrastructures. \\

In summary, the contributions of this paper are as follows:
(i) A conceptual testbed architecture that facilitates dual self-integration of various decentralized services by an IoT application, and different IoT applications by a decentralized service. 
(ii) The realization of the conceptual testbed architecture by abstracting the software engineering complexity in two software agents and their interactions in a generic communication protocol.
(iii) An improved and extended distributed prototyping toolkit for decentralized socio-technical systems of TRL-6.
(iv) Improvements of the I-EPOS software artifact~\cite{pournaras2018decentralized} that transitions from simulations to live deployment with a demonstrated TRL-6 continuous operation. 
(v) A proof of concept based on the self-integration and experimental evaluation of two decentralized services for IoT applications under highly dynamic environments. 

The rest of this paper is outlined as follows: 
Section~\ref{S:RW} reviews relevant previous work. 
Section~\ref{S:Arch} introduces the testbed architecture, and the realization protocol.
Section~\ref{S:DISC} illustrates the \hlc{Livepeer} toolkit, and Section~\ref{S:SS} introduces the two studied services.
Sections~\ref{S:ExpMeth} and~\ref{S:ExpEval} illustrate the experimental methodology, and evaluations, respectively.
Finally, Section~\ref{S:Conclusion} concludes this paper and outlines future work.
%%%%%
%%%%%
%%%%%
\begin{table*}[!htb]
\centering
\caption{Comparison of related work.}
\caption*{
\textbf{Symbols:} PT = Physical Testbed. PaaS = Platform-as-a-Service, TaaS = Testbed-as-a-Service, AO-Arch = Agent-Oriented Architecture. H = Hardware, D = Device, S = Service, Te = Technical, Sy = Syntactical, Se = Semantical.
}
\scriptsize
\begin{tabular}{p{21mm}p{13mm}p{14mm}p{12mm}p{6mm}p{9mm}p{11mm}p{12mm}p{17mm}} 
\toprule
\textbf{Related Work} &\textbf{Paradigm} & \textbf{Abstraction} & \multicolumn{4}{c}{\textbf{Socio-Technical System Design Considerations}} & \textbf{Reusability} & \textbf{Interoperability}\\\cmidrule{4-7}
& & &  Data \newline Locality & Privacy & Autonomy & Decentralized Control & \\
\addlinespace
\toprule
FIT IoT-Lab~\cite{adjih2015fit} & PT & H & - & - & - & - & H & Te + Sy \\
\addlinespace
SmartSantander~\cite{sanchez2014smartsantander} & PT & H & - & \checkmark & - & - & H & Te + Sy \\
\addlinespace
City of Things~\cite{latre2016city} & PT & H  & - & - & - & - & H & Te  \\
\addlinespace
CityLab~\cite{struye2018citylab} & PT & H & - & - & - & - & H & Te\\
\addlinespace
SmartCampus~\cite{nati2013smartcampus} & PT & H & - & - & - & - & H & Te + Sy \\
\addlinespace
MakeSense~\cite{jiang2019makesense} & PT & H & - & \checkmark & - & - & H  & Te   \\ 
\addlinespace
VICINITY~\cite{vicinity} &  PaaS  & H + D & - & \checkmark & - & - & H + D & Te + Sy + Se  \\
\addlinespace
IoTbed~\cite{hossain2017iotbed} & TaaS  & H & - & - & - & - & H & Te + Sy  \\
\addlinespace
Xively~\cite{sinha2015xively} & PaaS  & D  & - & - & - & - & D & Sy  \\
\addlinespace
Lysis~\cite{girau2016lysis} &  AO-PaaS & D (AO) & \checkmark & \checkmark & \checkmark & \checkmark & D & Sy + Se  \\
\addlinespace
AoT~\cite{mzahm2013agents} & AO-Arch  & D (AO) & - & - & \checkmark & - & D & Te + Sy  \\
\addlinespace
SIoT~\cite{atzori2011siot} & AO-Arch  & D (AO) & - & - & \checkmark & \checkmark & - & Se   \\
\addlinespace
iSapiens~\cite{cicirelli2016isapiens} & AO-Arch  & D (AO) & - & - & \checkmark & - & D & Te + Sy   \\
\addlinespace
BEMOSS~\cite{pipattanasomporn2015bemoss} & AO-Arch & D (AO) & - & - & - & - & D & Te + Sy   \\
\addlinespace
UBIWARE~\cite{scuturici2012ubiware} & AO-Arch & D (AO) & - & - & - & - & D & Te + Sy + Se   \\ 
\addlinespace 
FIoT~\cite{do2017fiot} & AO-Arch & D (AO) & - & - & \checkmark & \checkmark & D & Te + Sy   \\
\addlinespace
ACOSO-Meth~\cite{fortino2017agent} & AO-Arch & D (AO) & \checkmark & - & \checkmark & - & D + S & Sy + Se    \\
\addlinespace
VIVO~\cite{luceri2018vivo} & Framework & D & \checkmark & \checkmark & - & - & D & Sy   \\
\addlinespace
iCore~\cite{giaffreda2013icore} & Framework & D (AO) & - & \checkmark & \checkmark & - & D + S & Te + Sy + Se    \\
\addlinespace
Fluidware~\cite{zambonelli2019towards} & Framework & D & \checkmark & \checkmark & - & \checkmark & D & Sy + Se \\
\addlinespace
\textbf{Proposed} & AO-Arch & D (AO) + S  & \checkmark & \checkmark & \checkmark & \checkmark & D + S & Sy  \\
\bottomrule
\addlinespace
\end{tabular}
\label{T:RW}
\caption*{
The abstraction indicates the three possible levels of applied abstraction: 
H: Hardware abstraction by providing software routines to access the hardware via programming interfaces.
D: Device abstraction by having virtualized counterparts for each IoT device at the system-level. 
The AO indicates whether virtual counterpart is an agent.
Agents are networked software components that autonomously perform specific tasks on device/user behalf by interacting with other agents and with their environment~\cite{fortino2017agent}.
S: Indicates service-level abstraction by providing common communication protocols for IoT services.
Data locality, refers to local processing of data, \hlc{and} autonomy is the ability of the device to autonomously interact and actuate its function.
Decentralized control indicate the existence/lack of central control entities at the service-level.
Reusability refers to the ability to reuse the Hardwares (H), Devices (D), or Services (S) in different application scenarios.
The interoperability illustrates the utilized communication paradigm.
Technical refers to technological approaches (e.g., bluetooth), syntactical the shared message formats, and semantical the use of shared ontologies and knowledge representation~\cite{savaglio2020agent}.}
\end{table*}
%%%%%
%%%%%
%%%%%
\section{Related Work}
\label{S:RW}
%%%%%
%%%%%
%%%%%

\hlc{Self-adaptive frameworks have been earlier introduced to address the complexity, heterogeneity and uncertainties}~\cite{novoa2016self,krupitzer2015survey} \hlc{of large-scale integrated networked systems such as pervasive/ubiquitous computing and IoT}~\cite{baresi2010live}. \hlc{Such frameworks study adaptive service composition in dynamic environments at runtime, utilizing various techniques, such as context-aware computing}~\cite{urbieta2017adaptive,geihs2012context}, \hlc{service re-selection heuristics}~\cite{barakat2018adaptive} \hlc{and parallel service execution}~\cite{brun2013design,ferscha2015collective}. \hlc{However, these frameworks often do not address the self-integration of different physical devices at runtime}~\cite{novoa2016self}. \\

%\hlc{Due to their size, complexity, heterogeneity, large-scale integrated networked systems (e.g., pervasive/ubiquitous computing, IoT) pose new challenges regarding adapting to dynamic environments and integrating novel services and devices at runtime}~\cite{novoa2016self,krupitzer2015survey}.
%\hlc{In pervasive/ubiquitous computing, self-adaptive frameworks have been researched and utilized to address these challenges}~\cite{baresi2010live}.
%\hlc{Such frameworks study adaptive service composition in dynamic environments at runtime, utilizing various techniques, such as context-aware computing}~\cite{urbieta2017adaptive,geihs2012context}, \hlc{service re-selection heuristic}~\cite{barakat2018adaptive}, \hlc{and parallel service execution}~\cite{brun2013design,ferscha2015collective}.
%\hlc{However, these frameworks often do not address the self-integration of different physical devices at runtime}~\cite{novoa2016self}. \\

In the field of IoT, experimental facilities and physical testbeds have been subject to extensive previous research and surveys~\cite{arasteh2016iot,chernyshev2017internet,gluhak2011survey,savaglio2020agent,yr2019extensive,sotiriadis2014towards}.
Physical testbeds equip researchers with deployed and ready-to-use physical devices, simplifying the design and evaluation of novel IoT systems and services (e.g., network protocols, Big Data algorithms, city-wide IoT services) under realistic operational conditions~\cite{adjih2015fit,sanchez2014smartsantander,latre2016city,struye2018citylab,nati2013smartcampus,jiang2019makesense}.
One example is SmartSantander~\cite{sanchez2014smartsantander} with a city-wide scale ($\sim$20,000 sensors).
SmartSantander nodes act only as data sources and can be configured (centrally via a management plane) to run applications such as environmental monitoring.
However, in physical testbeds, often the domain/project-specific requirements determine the design and technological aspects (i.e., communication protocols) of sensors, smart objects, and middleware.
This limits their reusability in different domains and applications. 
To tackle such challenges, the \textit{PaaS} (platform-as-a-service) model has been studied and utilized.
The PaaS model leverages standard interfaces and interoperability measures, to provide researchers with tools to rapidly develop, execute, and manage IoT systems without the complexity of building and maintaining the infrastructure~\cite{girau2016lysis}.
This enables the design and deployment of cross-application IoT platforms~\cite{girau2016lysis}.
Xively\footnote{\href{https://xively.com}{https://xively.com [Last accessed: May 2020]}} is an example in the context of distributed cloud-based applications with a centralized control plane, where different tasks are executed in separate platforms and devices. 
For instance, application-level functions can be executed in different virtual and real entities to reduce latency and bottlenecks.
Nevertheless, PaaS approaches often neglect socio-technical requirements, such as data locality, privacy, autonomy, and decentralized control. \\

Agent-based computing has been used extensively to enable cooperative, decentralized, dynamic, and open IoT systems~\cite{savaglio2020agent}.
In such systems, agents autonomously interact and cooperate based on (typically) asynchronous message passing mechanisms to perform a task or a service.
Shared communication standards facilitate agent interoperability and allow for incorporating heterogeneous resources.
Examples of such systems include Lysis~\cite{girau2016lysis}, which introduces a PaaS model with virtualized autonomous social agents, allowing for the deployment of fully distributed applications.
ACOSO-Meth~\cite{fortino2017agent} introduces an agent-oriented architecture based on IoT smart objects, as well as a taxonomy for assessing system-level requirements and technological readiness of IoT systems.
While agent-based approaches utilize device virtualization and address some socio-technical considerations, on the service-level they often suffer from lack of standard interfaces and interoperability.
Thus, to reuse a specific service in different applications, the code and communication protocol should change.
The proposed architecture in this paper utilizes service abstraction to enable the reusability of devices and services in various application domains. \\

IoT systems are operated in increasingly dynamic and complex environments~\cite{zambonelli2019towards}, where during system runtime, devices can fail, users might join/leave, communication across devices and agents becomes disrupted, system goals and requirement can vary, and new services are required.
Not all such changes can be foreseen during the initial design phase.
Often it is infeasible for a centralized controller to have knowledge of all such changes in a timely manner.
\hlc{Self-adaptive} approaches, autonomic computing~\cite{kephart2003vision,landauer2015designing}, and hierarchical self-aware decision-making~\cite{diaconescu2018hierarchical} have been studied as means to handle such changes at runtime, with minimal human intervention~\cite{tomforde2016organic,barnes2019chariot,bellman2019self,landauer2015designing}.
To this end, the proposed testbed architecture facilitates the rapid prototyping and experimentation of IoT services that can handle dynamic environments (with high TRL) as well as autonomously initialize and include various devices and services during runtime. 
Table~\ref{T:RW} illustrates a non-exhaustive comparison between relevant previous research, providing insights of the existing experimental IoT testbeds\footnote{Note that the comparisons and distinction of the socio-technical considerations are based on the system design goals rather than subsequent third-party augmentations and applications.}.
%%%%%
%%%%%
%%%%%
\section{A Conceptual IoT Testbed Architecture for System Self-integration}
\label{S:Arch}
%%%%%
%%%%%
%%%%%
\begin{figure*}[!htb]
    \centering
    \subfloat[Monitoring power demand and optimizing traffic.]{\includegraphics[width = 0.51\textwidth]{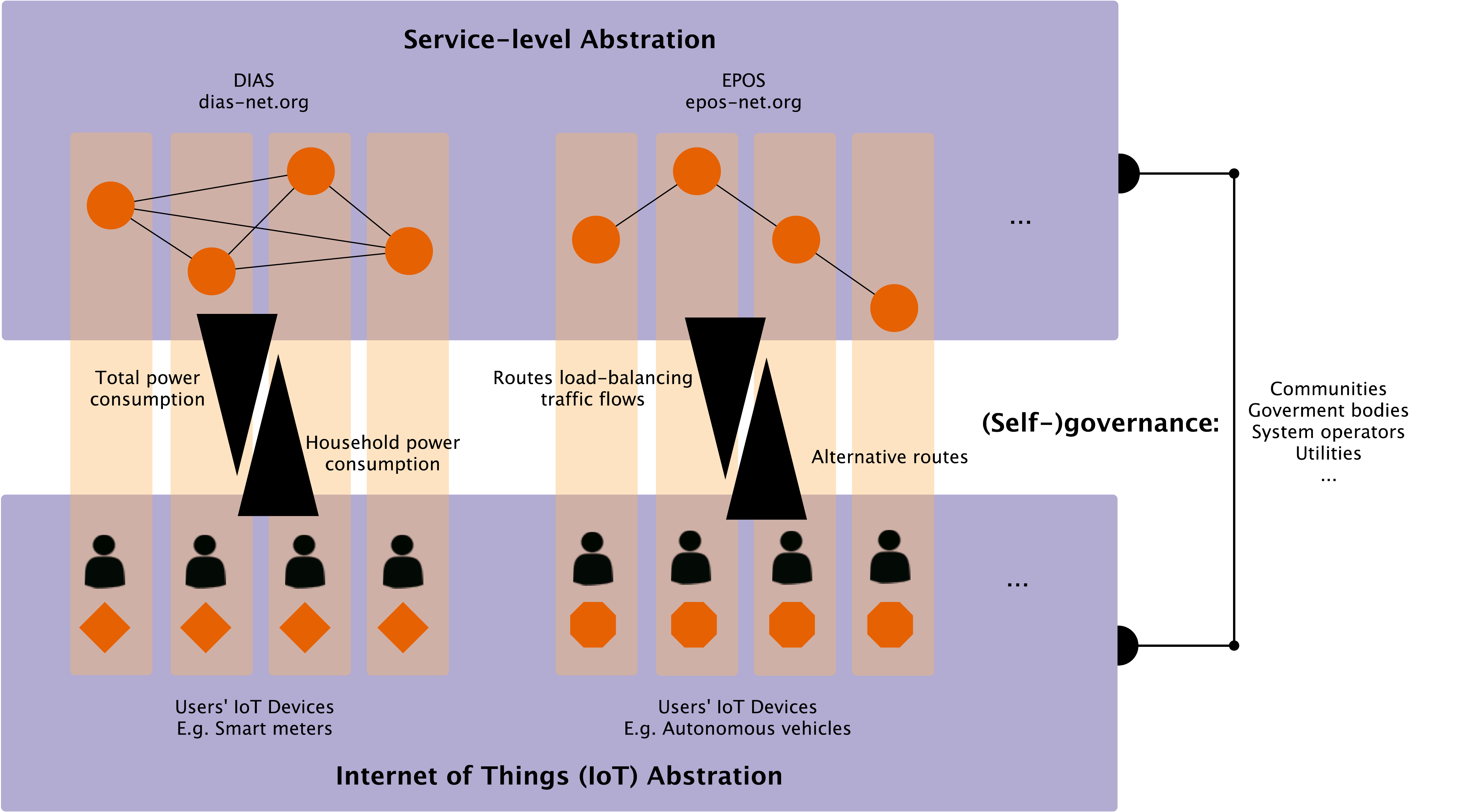}}
        \subfloat[Monitoring traffic and optimizing power demand.]{\includegraphics[width = 0.51\textwidth]{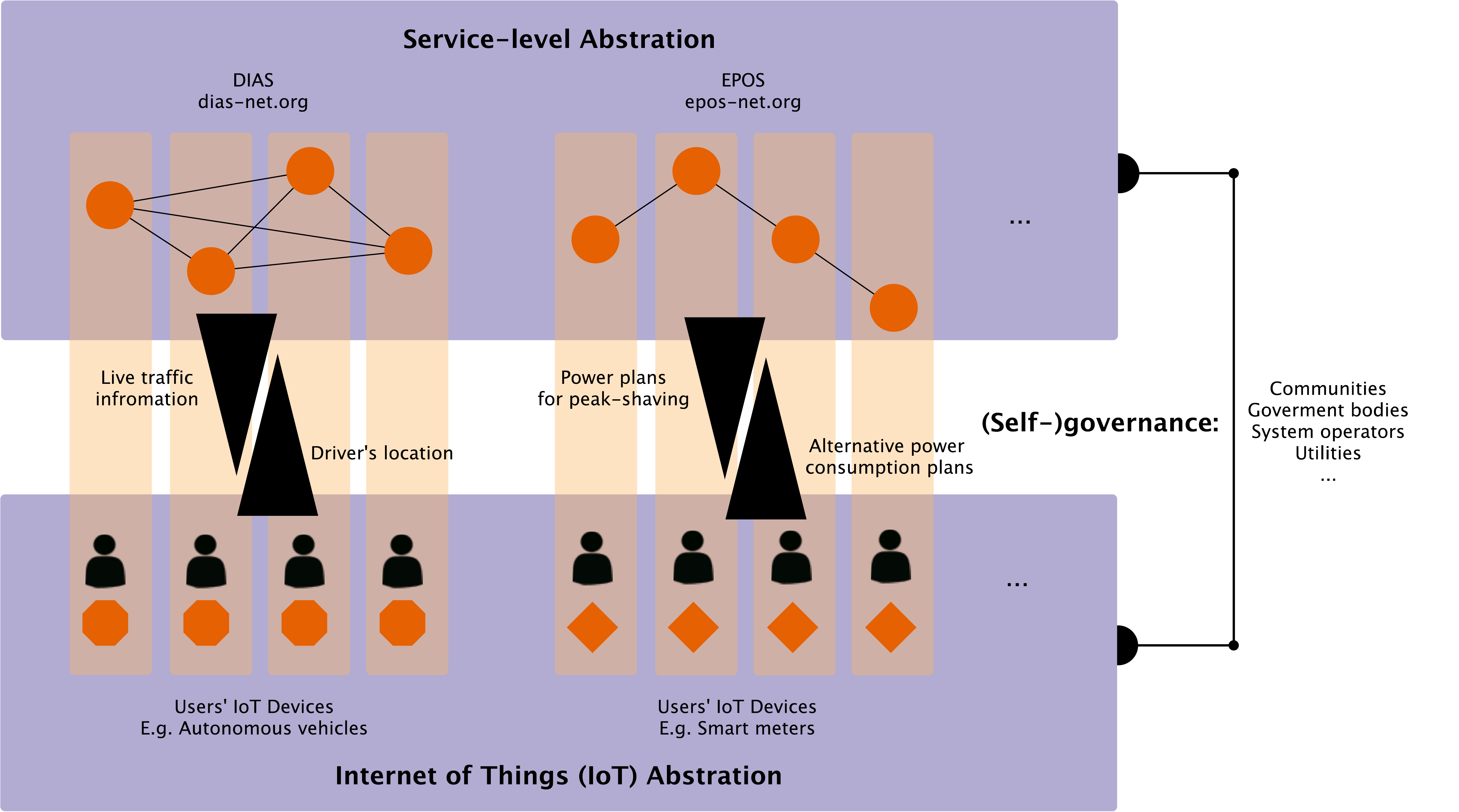}}
    \caption{Conceptual testbed architecture and two examples of application scenarios by self-integrating different decentralized services with different IoT applications. 
Note how by switching the coupling of the two IoT applications with the two decentralized services, new application scenarios are seamlessly supported.}
\label{fig:architecture}
\end{figure*}
%%%%%
%%%%%''[
%%%%%
This paper introduces \hlc{a} conceptual testbed architecture, designed to enable seamless reusability and self-integration of (i) application-independent decentralized services by an IoT application, and (ii) different IoT applications by the same decentralized service. 
\hlc{This architecture focuses on decentralized socio-technical IoT services with autonomous agents, without central authority to coordinate the agents and their actions.
These agents interact locally with their users and remotely with each other to make intelligent collective choices, via which they can self-regulate the consumption and production of common resources.
Hence, in this context a service is essentially a distributed software running on multiple agents.}
Examples of such services include monitoring services~\cite{kelly2013towards}, real-time analytics~\cite{pournaras2017engineering}, planning and coordination systems~\cite{pournaras2018decentralized}, learning techniques~\cite{shanthamallu2017brief}, and distributed control systems~\cite{pournaras2017self}.
Figure~\ref{fig:architecture} illustrates the conceptual testbed architecture, and two examples of self-integrating different decentralized services with different IoT applications. \\

\hlc{To enable the aforementioned reusability and self-integration,} the proposed architecture utilizes two levels of abstraction:
(i) IoT application level, and (ii) decentralized service level.
At the IoT application level, this abstraction creates \emph{application agents}: a piece of software lying on each user's IoT device, acting as the middleware for communication with the decentralized service.
\hlc{These} IoT devices provide sensing and actuation capabilities. They are of different types (e.g., sensors, mobile phones) and geo-spatially distributed.
At the decentralized service level, this abstraction creates \textit{service agents, as one-to-one} counterparts for each application agent.
Each service agent has the following tasks: 
(i) Receiving data from the corresponding application agent (IoT device).
(ii) Executing the service by interacting and cooperating with other service agents.
Finally, (iii) Providing the outcome of the service to the \hlc{application agent} (e.g. in the form of control commands).
This dual abstraction creates a decoupling between the internal operations of the IoT application and the complex functionality of the decentralized service:
The first abstraction level (application to services) facilitates the inclusion of heterogeneous devices, and their reusability of their applications in different services, while the second abstraction level (service to applications) simplifies the reusability of decentralized services in new applications, as the interfaces, communication logic, and protocols remain unaffected by changes in the IoT devices or applications.
\hlc{In production-ready systems, both the application and service agents can run on the same computational node to reduce latency and provide data locality, i.e. on a user's device such as a smartphone, or at two remote nodes, i.e. on a user's device, a cloud node, or a crowdsourced community server}\footnote{Such as the Diaspora~\cite{bielenberg2012growth} foundation: \href{diasporafoundation.org}{diasporafoundation.org} [Last accessed: May 2020]}. \\

The deployment and governance of the testbed depend on the \hlc{target IoT application.}
The service agents are deployed and managed by the service operator, which can be a third-party mediator in sensing-as-a-service scenarios~\cite{yr2019extensive}, or a community in case of participatory sensing applications~\cite{tilak2013real}.
Examples of third-party mediators include companies such as Waze\footnote{\href{https://www.waze.com}{https://www.waze.com} [Last accessed: May 2020]}, Uber\footnote{\href{https://www.uber.com}{https://www.uber.com} [Last accessed: May 2020]}, and Swiss Mobility\footnote{\href{https://www.mobility.ch}{https://www.mobility.ch} [Last accessed: May 2020]}, while environmental monitoring~\cite{shah2016iot}, and urban sensing~\cite{calabrese2015urban} are examples of participatory sensing applications deployed and managed by service communities.
Furthermore, Smart City scenarios run by municipalities~\cite{medvedev2015waste}, Smart Grids~\cite{collier2016emerging}, and smart supply chains~\cite{wu2016smart} are examples where a central authority such as the municipality or the utility company governs the system.
However, in participatory sensing scenarios, such as environmental monitoring~\cite{shah2016iot}, and urban sensing~\cite{calabrese2015urban}, the testbed can be self-governed by users and the service community (public good infrastructure).
%%%%%
%%%%%
%%%%%
\begin{table}[t]
\centering
\caption{\hlc{System entities}}
\resizebox{\columnwidth}{!}{%
\begin{tabular}{l l l} 
\toprule
\textbf{Entity} & \textbf{Explanation} & \textbf{Example (Figure~\ref{fig:architecture}a)} \\
\midrule
IoT \newline Application & Control logic & Monitoring power demand \\
\addlinespace
IoT Device & Sensing and actuation & Smart meters \\
\addlinespace
IoT Service &  Autonomous general-purpose agents & Collective measurements \\
\addlinespace
Gateway &  System bootstrapping proxy for service agents & Software \\
\addlinespace
Service Operator & Setting up service agents and the gateway & Communities, power utility \\
\bottomrule
\end{tabular}
}
\label{T:terms}
\end{table}
%%%%%
%%%%%
%%%%%
\subsection{Communication Protocol \& Runtime Cycle}
\label{S:RP}
%%%%%
%%%%%
%%%%%
\begin{figure}[!htb]
\centering
\includegraphics[width = 0.5\textwidth]{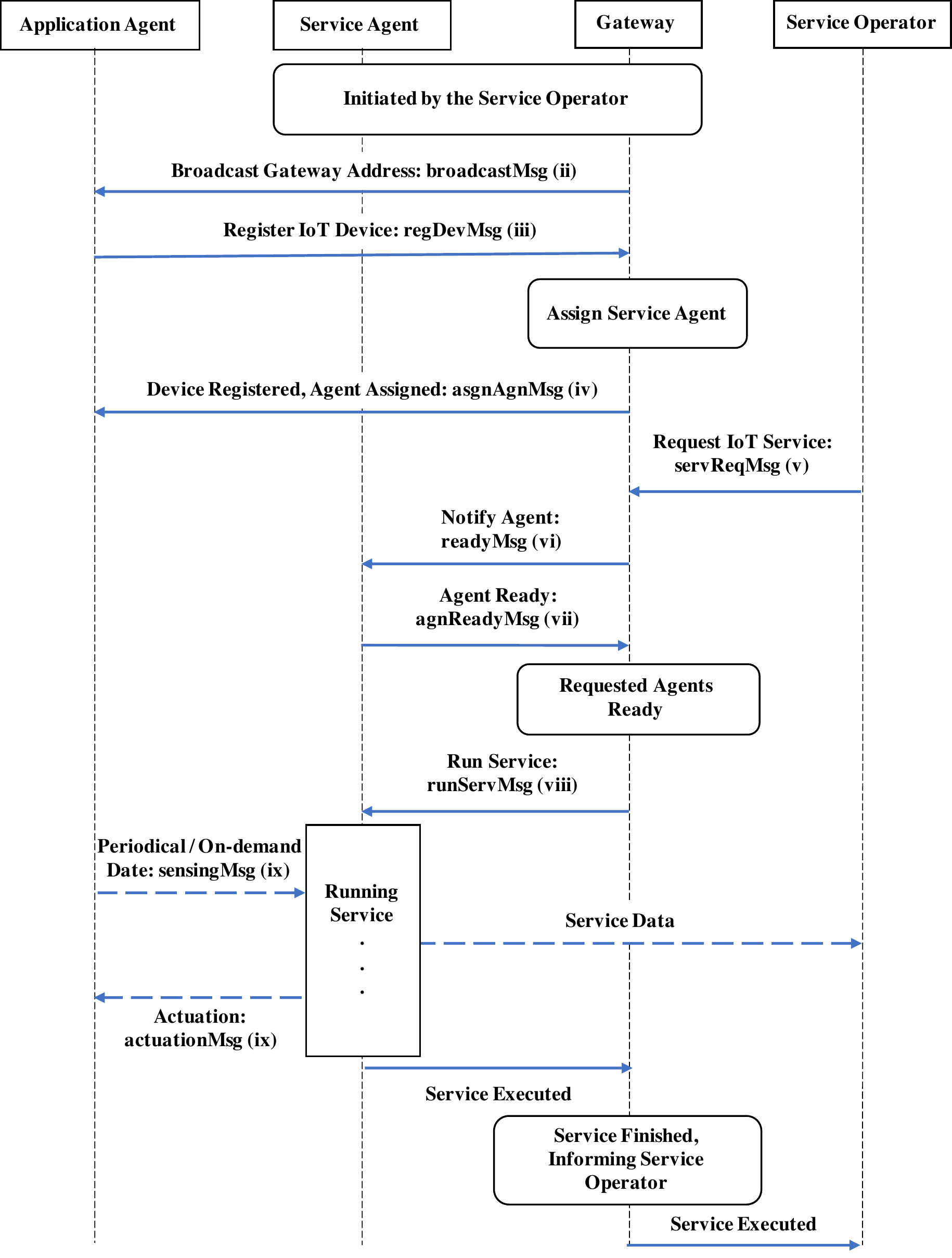}
\caption{The communication protocol that realizes the conceptual testbed architecture.
The details of messages are illustrated in Section~\ref{S:Arch}.
This protocol treats the running services as blackboxes and create standard interface between different components of the testbed.
Hence, different devices and services can be self-integrated at runtime.}
\label{fig:runtimeCycle}
\end{figure}
%%%%%
%%%%%
%%%%%
\hlc{A distributed protocol is designed for the communication and self-integration between the two abstracted levels}.
This generic communication protocol is application and service independent. \hlc{It determines the communication logic and common interfaces between service and application agents.}
Table~\ref{T:terms} \hlc{illustrates the various system entities in the protocol according to }Figure~\ref{fig:architecture}a.
Figures \ref{fig:runtimeCycle} illustrates the protocol sequence diagram, and runtime cycle.  
The protocol is outlined as follows:
(i) The service operator initializes the service agents, and the gateway.
The gateway act as a bootstrapping proxy, \hlc{connecting the service agents} to the corresponding application agent.
The gateway is agnostic of data, the internal processes of IoT applications and decentralized services\footnote{In practice, the gateway does not need to be a separate entity, and can be incorporated in a service agent.}.
(ii) It is assumed that application agents know the public address of the gateway.
This is possible via the \verb|broadcastMsg:{GWAddr, servInfo}| by the gateway, where \verb|GWAddr| is the gateway address, and \verb|servInfo| indicates the service.
(iii) To connect to the decentralized IoT service, each application agent contacts the service gateway, registers itself, and its corresponding IoT device, via the \verb|regDevMsg:{devAddr, devInfo, servInfo}|, where \verb|devAddr| is the application agent's address, and \verb|devInfo| \hlc{includes information such as device type, and location.}
(iv) In response, the gateway assigns a service agent to the IoT device and informs the application agent via the \verb|asgnAgnMsg:{agnAddr}|, where \verb|agnAddr| is the address of the assigned service agent.
(v) The service operator submits a service request to the gateway, specifying its requested service, and execution metadata, via the \verb|servReqMsg:{servInfo, servMD}|, where \verb|servMD| contains the metadata required to execute the service, such as the number of service agents, number of devices, and their locations.
(vi) The gateway receives the service request, and notifies the service agents, via the \verb|readyMsg:{servInfo, servMD}|.
(vii) The service agents validate the service information \hlc{along with the associated metadata,} and reply to the gateway, via the \verb|agnReadyMsg:{agnAddr, servInfo}|.
(viii) When all agents are notified and ready to run the service, the gateway sends the execute service command, via the \verb|runServMsg:{servInfo}|.
(ix) \hlc{Each} service agent requests/receives data from the application agent via the \verb|sensingMsg:{servInfo, data}|, and submits control/actuation messages to the IoT device either periodically, on-demand, or at the end of the service execution via the \verb|actuationMsg:{servInfo, actuation}|.
(x) Finally, after the service is executed, the application agent, gateway, and the service operator are informed.
%%%%%
%%%%%
%%%%%
\section{Prototyping Decentralized IoT Systems}
\label{S:DISC}
%%%%%
%%%%%
%%%%%
%\hlc{To realize the introduced conceptual architecture, and facilitate the prototyping of IoT applications and services, several challenges should be addressed.} 
\hlc{To ensure efficient and reliable performance, continuous testing and refinements are needed, from the early stages of simulation, to live deployment, to long-lasting stable operation.
Moreover, high costs and complexity are involved in the maintenance of different implementations when code is changed back and forth to validate new functionality and expand to new application domains.}
To address these challenges, this paper introduces the \hlc{Livepeer} toolkit.
\hlc{Livepeer} is based on an improved version of the general-purpose prototyping toolkit Protopeer~\cite{galuba2009protopeer}, now made highly robust and efficient for long-term operations.
\hlc{Livepeer comes with a high Technical Readiness Level (TRL-6) and its new modules include:}
(i) \hlc{The redesigned and reengineered Protopeer node, providing core functionality such as communication protocols (e.g. TCP messaging), timers, and an execution environment for the service agents.}
(ii) Software clients, \hlc{acting as application agents,} for supporting IoT devices (i.e. smartphones). 
(iii) A networking module \hlc{for efficient and reliable application-to-services and service-to-applications communication}. 
(iv) A scalable monitoring infrastructure, integrated to each computational node, for application/service monitoring and analysis.
%%%%%
%%%%%
%%%%%
\subsection{Livepeer: Redesigning \& Reengineering Protopeer}
\label{S:proto}
%%%%%
%%%%%
%%%%%
The Protopeer toolkit~\cite{galuba2009protopeer} is designed with the main goal of facilitating the rapid prototyping of P2P applications, and the transition from simulation to live environments.
\hlc{However, the transition from live environments to robust, long-lasting, TRL-6 operations is not trivial, as the latter often has a larger scale, higher realism regarding performance degradation by network partitions, and message losses.
For instance, long-term operations require magnitude resources, incur a higher number of operations, and communicate a larger volumes messages.
This results in the discovery of unforeseen system faults and deficiencies, such as sporadic message losses, deadlocks, synchronization, excessive thread counts, and memory leaks by evolving communication processes.
To address the above challenges, this paper introduces a redesigned and reengineered version of Protopeer for highly efficient and robust long-lasting experimentation.
In addition to several implementation-specific improvements, the redesigned Protopeer includes: 
(i) A redesigned communication and networking module, enabling multiple queues for message processing} (Section~\ref{S:CP}), \hlc{and
(ii) a novel module for distributed monitoring and logging of services, events, and memory usage} (Section~\ref{S:LMI}). 
This new system \hlc{has} been tested extensively over several weeks under highly dynamic environments, with approximately 3000 node join/leaves, 150,000 runtime parameter changes, and over 2.1 million exchanged messages per day (Section~\ref{S:EvalScenII}). \\
%%%%%
%%%%%
%%%%%
\begin{figure}[t]
\centering
\includegraphics[width = 0.45\textwidth]{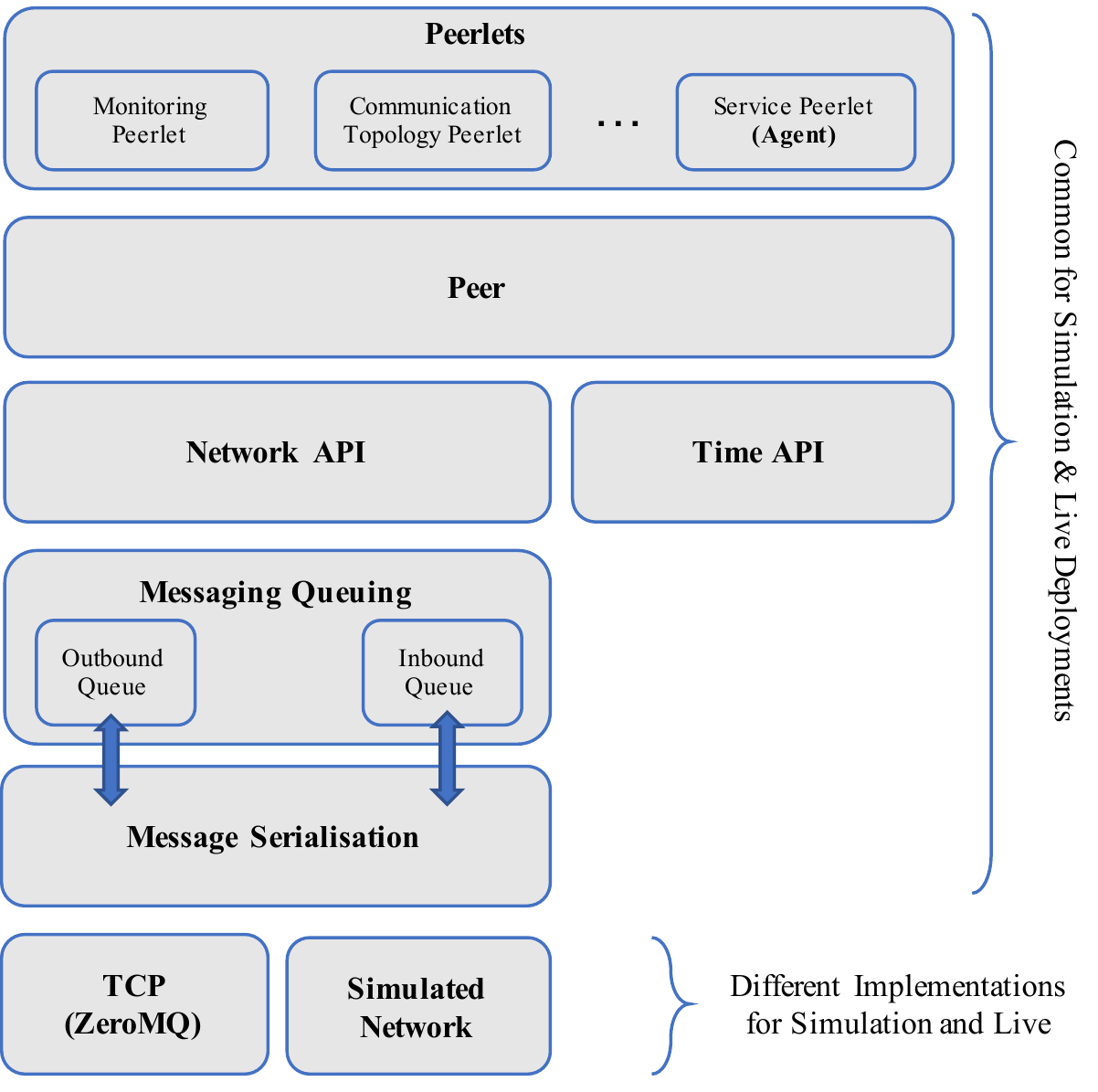}
\caption{Internal architecture and modules of the \hlc{Livepeer} node, a redesigned version of the Protopeer node.
The modules are separated in two categories: deployment-dependent and deployment-independent.
This design has the advantage to first test a system in a controlled simulated environment and then gradually move to a large-scale live deployment, while the deployment-dependent modules are the only ones that require change.}
\label{fig:ProtopeerConcise}
\end{figure}
%%%%%
%%%%%
%%%%%

Figure~\ref{fig:ProtopeerConcise} illustrates a summarized view of the internal architecture of a Livepeer \hlc{node.}
There are two core concepts within each node: the peer, and the peerlet.
The peer provides core functionality such as communication protocols (e.g., TCP messaging), and timers. It acts as the execution environment (container) for the peerlets.
The peerlets, on the other hand, are independent modules that provide specific functionality and tasks.
Typically a node consists of a single Livepeer peer and multiple peerlets that collectively fulfill the required functionality of the service.
\hlc{In addition to the service agent,} two other examples of peerlets are the communication topology peerlet, and the monitoring peerlet.
The communication topology peerlet determines the service network topology and the communication logic, 
for instance, a tree-topology~\cite{pournaras2014adaptive} \hlc{(utilized in I-EPOS, see} Section~\ref{S:IEPOS}), or the gossip-based peer sampling for P2P systems~\cite{jelasity2004peer} \hlc{(utilized in DIAS, see} Section~\ref{S:DIAS})
The monitoring peerlet stores and submits logs from different modules in the peer to the monitoring infrastructure (Section~\ref{S:LMI}).
By default, this peerlet includes three different logging modes: 
(i) \emph{Service logger}, which logs the specific service logs.
(ii) \emph{Event logger}, which provides event-based logging, and insights into execution sequence.
(iii) \emph{Memory logger}, which measures the total memory footprint of a peer in memory, including nested objects.
%%%%%
%%%%%
%%%%%
\subsection{Communication Platform}
\label{S:CP}
%%%%%
%%%%%
%%%%%
\hlc{The proposed testbed} utilizes a messaging protocol based on a fast and lightweight TCP/IP implementation using ZeroMQ for agent-to-agent communication.
Each agent is instantiated with a single PULL socket and multiple PUSH sockets.
Thus, each agent can act as a sink and receives messages from any other agents in the network, whilst simultaneously sending messages to other agents through the PUSH sockets.
This is achieved by implementing two independent messaging queues, one for inbound messages, and one for outbound messages.
This separation allows the monitoring of queue size within each agent to regulate traffic flows.
%%%%%
%%%%%
%%%%%
\subsection{Monitoring Infrastructure}
\label{S:LMI}
%%%%%
%%%%%
%%%%%
A major challenge in decentralized IoT services with autonomous agents is to log and monitor the service, both at the individual agent level, as well as system-wide~\cite{wang2018fear}.
Devices and agents can be geo-spatially distributed, and deployed over different computational clusters and networks~\cite{bonomi2012fog}.
While each agent can run autonomously and independently, most analytics require an aggregated view in real-time, whilst also allowing to drill-down and investigate the internal activities within a single agent.
Providing such multi-granular views is even more challenging for large-scale systems~\cite{cai2016iot}.
Simple NFS (Networked File System) solutions may not provide the necessary throughput to sustain heavy logging from several agents~\cite{cai2016iot}.
Additionally, IoT devices and agents can be restricted in computational resources, hence, the logging and monitoring needs to be lightweight, simple to use, efficient, and with minimal impact on the real-time performance~\cite{trihinas2018low}. \\

To address these challenges, this paper devises a monitoring infrastructure, comprised of the following components: 
(i) A single database, containing \hlc{the logged} data from agents.
(ii) A single logging gateway, accessible to all agents, which receives the logs and commits them to the database.
The main task of the logging gateway is to perform authentication, authorization, and connection pooling for the database.
(iii) A single, lightweight peerlet on each agent, known as the monitoring peerlet, which collects the logs and submits them to the logging gateway\footnote{The earlier version of the Protopeer toolkit saves logging objects locally that creates a discrepancy for post-processing and analysis as systems operate in the long run}.
This infrastructure is easy to integrate (by only adding the peerlet) in Livepeer nodes.
It is distributed, and modular in design, and can be connected to various observability platforms and dashboards, such as Grafana\footnote{\href{https://grafana.com}{https://grafana.com [Last accessed: May 2020]}} and Redash\footnote{\href{https://redash.io}{https://redash.io [Last accessed: May 2020]}} for real-time visualization.
%%%%%
%%%%%
%%%%%
\section{Studied Services}
\label{S:SS}
%%%%%
%%%%%
%%%%%
This paper studies two live implementations of generic multi-purpose IoT services as proof-of-concept and use-cases of the proposed architecture;
Namely, collective learning based on the I-EPOS system (Section \ref{S:IEPOS}), and decentralized collective measurements based on the DIAS system (Section \ref{S:DIAS}).
I-EPOS performs collective learning for multi-agent combinatorial problems~\cite{pournaras2018decentralized}, \hlc{utilizing a structured network topology (tree-topology) with synchronous learning iterations and communication between agents.}
DIAS performs decentralized privacy-preserving data analytics, \hlc{relying} on local computations, peer-to-peer interactions, and hashed information~\cite{pournaras2017engineering,pournaras2017self}.
It has an unstructured topology (P2P) for asynchronous communication~\cite{pournaras2017engineering}.
Due to their decentralized socio-technical design, both of these services are very relevant to IoT applications.
However, they are profoundly different in their operation, which challenges the flexibility and applicability of the proposed architecture.
%%%%%
%%%%%
%%%%%
\subsection{Collective Learning}
\label{S:IEPOS}
%%%%%
%%%%%
%%%%%
The I-EPOS system~\cite{pournaras2018decentralized} performs fully decentralized, self-organizing, privacy-preserving combinatorial optimization\footnote{Available at http://epos-net.org [Last accessed: May 2020]}.
The I-EPOS agents (service agents) provide the learning service, and each has a set of local plans generated by the application agent.
These plans can be alternative routes from an autonomous vehicle, or power consumption schedules from a smart appliance (e.g., smart washing machine).
I-EPOS optimizes a system-wide goal, measured by a global cost function.
This goal can be load-balancing traffic flows in a city~\cite{gerostathopoulos2019trapped}, or peak-shaving power demand for Smart Grids~\cite{fanitabasi2019appliance}.
The I-EPOS agents interact and cooperate with each other to select a plan that minimizes the global cost.
The agents self-organize in a tree-topology~\cite{pournaras2014adaptive} as a way of structuring their interactions.
I-EPOS performs consecutive learning iterations, which includes two phases: the \emph{bottom-up} (leaves to root) phase and \emph{top-down} (root to leaves) phase. 
At each iteration $t$, agent $u$ selects the plan $p_{u,s}^t$ to satisfy the following optimization objective:
%%%%
%%%%
%%%%
	\begin{equation}
	\begin{aligned}
	\label{eq:IEPOSEConcise}
	p_{u,s}^t := \arg\min_{j=1}^{|P_u|}~\Bigg((1-(\alpha+\beta))~\Big(f_G(p_{u,j}^t)\Big) +\\
	\beta~\Big(f_L(p_{u,j}^t)\Big) + \alpha~\Big(f_U(p_{u,j}^t)\Big)\Bigg)
	\end{aligned}
	\end{equation}
%%%%
%%%%
%%%%
In the above equations, $|P_u|$ is the number of plans for agent $u$, $f_G(p_{u,j}^t)$ is the global cost of selecting $p_{u,j}^t$, which can be the variance of traffic load across different routes (in case of traffic load-balancing).
Each plan has a local cost, calculated by $f_L(p_{u,j}^t)$, which can be the trip duration (in case of alternative routes), or user discomfort (in case of shifting power consumption).
$f_U(p_{u,j}^t)$ is the unfairness, calculated by the dispersion of the local cost of the selected plans over all agents, with lower values indicating more equal distribution of the local cost across all agents.
The $\alpha$, $\beta$, and $1-(\alpha+\beta)$ parameters indicate the agents' preferences for unfairness, local cost, and global cost, respectively.
For instance, an agent with $\alpha=0,~\beta=1$ is known as a \textit{selfish} agents, which prioritizes the minimization of its local cost, while another agent with $\alpha,\beta=0$ is known as an \textit{altruistic} agent, which minimizes the global cost.
After the final iteration $F$ is completed, $p_{u,s}^F$ is presented to the users' devices for execution.
Further elaboration on I-EPOS is out of the scope of this paper and the interested reader is referred to previous work \cite{pournaras2018decentralized}.
%%%%%
%%%%%
%%%%%
\subsection{Decentralized Collective Measurements}
\label{S:DIAS}
%%%%%
%%%%%
%%%%%
The DIAS system~\cite{pournaras2017engineering} performs fully decentralized privacy-preserving data analytics for the Internet of Things\footnote{Available at http://dias-net.org [last accessed: May 2020]}.
Each application agent acts as a data supplier and consumer:
Data suppliers are sensors that locally generate a stream of real-time privacy-sensitive data, while data consumers collect these data to compute information (e.g., summation, average, max/min, top-k.).
For instance, data suppliers provide consumption data from residential smart meters, and data consumers receive the aggregated power consumption of the neighborhood, as well as status updates about the reliability of the Smart Grid~\cite{pournaras2014decentralized}.
This process is fully decentralized and privacy-preserving, as users' data is not shared with a central entity.
Each pair of data supplier and consumer is connected to a DIAS agent (service agent).
Each data supplier discovers data consumers in the service network, to which the local sensor data is sent.
This discovery is performed via a fully decentralized gossiping protocol: the peer sampling service~\cite{jelasity2007gossip}.
Data suppliers spread the local sensor data in the network periodically by pushing them to remote data consumers to maintain a high accuracy in the estimations of the aggregation functions.
Each data consumer collects the input data for the computation of the aggregation functions. 
Finally, data suppliers receive the outcome of the performed aggregation. 
Each bilateral interaction between a data supplier and consumer is referred to as an aggregation session. \\

The raw data from data suppliers can be privacy-sensitive and rapidly varying over time.
To tackle this challenge, DIAS utilizes data summarization by assigning the raw values from stream data to a selected state chosen from a limited number of $k$ possible states.
\hlc{Additionally, DIAS addresses two other uncertainties: changes in the set of possible states, and agents leaving/failing/rejoining the network.
The challenge here is to preserve the accuracy of DIAS estimations under these two dynamics.}
To address this, DIAS uses a distributed memory system based on Bloom filters~\cite{bloom1970space} to track the history of the performed computations and when needed perform self-corrective actions.
Further elaboration on DIAS is out of the scope of this paper and the interested reader is referred to previous work~\cite{pournaras2017self,pournaras2017engineering}.
%%%%%
%%%%%
%%%%%
\section{Experimental Methodology \& Settings}
\label{S:ExpMeth}
%%%%%
%%%%%
%%%%%
The experiments in this paper are divided into two evaluation scenarios, both utilize the \hlc{Livepeer} toolkit, and follow the conceptual architecture, and communication protocol illustrated on Figures~\ref{fig:architecture} and~\ref{fig:runtimeCycle}, respectively.
The first evaluation scenario (Section~\ref{S:MethScenI}) studies the accuracy of the two services in live environments, to provide a performance benchmark for the testbed architecture and \hlc{Livepeer} toolkit given experimental realism~\cite{sanchez2014smartsantander}.
The second evaluation scenario (Section~\ref{S:MethScenII}) studies the efficiency and robustness of the services during long-lasting operation under dynamic and volatile environments.
%%%%%
%%%%%
%%%%%
\subsection{Evaluation Scenario I: Comparing Accuracy in Non-volatile Environments}
\label{S:MethScenI}
Experiments under live environments, even without dynamic changes, can incur inaccuracies due to networking errors (e.g., packet losses), clock differences across machines, and system failures~\cite{gluhak2011survey}.
This evaluation scenario analyzes the accuracy of the two studied scenarios, and provides a benchmark comparison in live non-volatile environments, to study the validity of the testbed architecture and the \hlc{Livepeer} toolkit.
For I-EPOS, this evaluation is made by comparing the simulation and live deployments of the service, and for DIAS the evaluation is made based on long-lasting operation with high experimental realism.
%%%%%
%%%%%
%%%%%
\subsubsection{I-EPOS} 
\label{S:MethScenIEPOS}
%%%%%
%%%%%
%%%%%
The experimental settings and parameters for I-EPOS in evaluation scenario I are illustrated in Table~\ref{T:IEPOSSettings}.
%%%%%
%%%%%
%%%%%
\begin{table}[t]
\centering
\caption{I-EPOS settings and parameters for the two evaluation scenarios.
The profiles for evaluation scenario I are illustrated in Table~\ref{T:ScenIProf} .
In evaluation scenario II, the number of agents range between $[150,250]$, and the $\alpha, \beta$ values range between $[0,1]$, given $\alpha+\beta =1$}
\scriptsize
\begin{tabular}{p{21mm}p{25mm}p{30mm}} 
\toprule
\textbf{Parameters} & \textbf{Value in \newline Evaluation Scenario I} & \textbf{Value in \newline Evaluation Scenario II}  \\
\midrule
Performed \newline Experiments & 100 per profile & Continuous: Intensity change every 8 hours \\
\addlinespace
Number of Agents & $50/100/300$ & $[150,250]$\\
\addlinespace
Dataset & EV Dataset & EV Dataset: 7-days ahead  \\
\addlinespace
Plans per Agent & 4 & 4 \\
\addlinespace
Plan Dimensions & $1440/4320/10080$ & 10080\\
\addlinespace
Number of Iterations & 50 & 50 \\
\addlinespace
Global Cost Function & MIN-VAR/RMSE & MIN-VAR/RMSE \\
\addlinespace
Local Cost Function & Discomfort & Discomfort \\
\addlinespace
Agent Preference & $\alpha=0$,$~~\beta=0/1$  & $\alpha, \beta \in [0,1],~\alpha+\beta=1$ \\
\addlinespace
Network Topology & Balance Binary Tree & Balance Binary Tree \\
\bottomrule
\end{tabular}
\label{T:IEPOSSettings}
\end{table}
%%%%%
%%%%%
%%%%%
\hlc{The utilized dataset contains charging plans for 2779 electric vehicles (EV) in three different planning horizons: 1, 3, and 7 days ahead}~\cite{pournaras2019socio}.
In all cases, each EV has 4 alternative charging plan in the form of a vector, specifying the energy demand for each minute during the planning horizon.
For 1-day-ahead plans, the length is 1440 ($24h*60min$), and for 3 and 7-days-ahead plans, the length is 4320 ($3d*24h*60min$), and 10080 ($7d*24h*60min$), respectively.
Two different global cost functions are applied to the total charging demand of the participating EVs, each addressing a different charging scenario:
(i) Minimizing charging demand variance (MIN-VAR), and
(ii) shifting charging times to night (MIN-RMSE)\footnote{MIN-RMSE: Minimizing the root mean square error between the total charging demand of all EVs, and the steering signal set by the service operator to incentivize night charging.
 The steering signal is a vector of the same length as the charging plans, with the day-time charging target set to 0.}.
Each plan also has a local cost, which is its discomfort calculated by the historical likelihood of using the EV while charging~\cite{pournaras2019socio}\footnote{Further elaboration on this dataset can be found in previous work~\cite{pournaras2019socio}.}.
%%%%%
%%%%%
%%%%%
The performed experiments are based on the 12 profiles illustrated in Table~\ref{T:ScenIProf}.
%%%%%
%%%%%
%%%%%
\begin{table}[t]
\centering
\caption{12 Profiles used for I-EPOS experiments scenario I}
\scriptsize
\begin{tabular}{p{7mm}p{13mm}p{15mm}p{14mm}p{16mm}} 
\toprule
\textbf{Profiles} & \textbf{Scale} & \textbf{Agent \newline Preference} & \textbf{Planning Horizon} & \textbf{Global Cost \newline Function} \\
\midrule
1 & & $\alpha=0$,$~~\beta=0$ & 1 day ahead  & MIN-VAR \\
\addlinespace
2 & Small & $\alpha=0$,$~~\beta=1$ & 1 day ahead  & MIN-VAR \\
\addlinespace
3 & (50 Agents)& $\alpha=0$,$~~\beta=0$ & 1 day ahead  & RMSE \\
\addlinespace
4 & & $\alpha=0$,$~~\beta=1$ & 1 day ahead  & RMSE \\
\addlinespace
\midrule
5 & & $\alpha=0$,$~~\beta=0$ & 3 days ahead  & MIN-VAR \\
\addlinespace
6 & Medium & $\alpha=0$,$~~\beta=1$ & 3 days ahead  & MIN-VAR \\
\addlinespace
7 & (100 Agents)& $\alpha=0$,$~~\beta=0$ & 3 days ahead  & RMSE \\
\addlinespace
8 & & $\alpha=0$,$~~\beta=1$ & 3 days ahead  & RMSE \\
\addlinespace
\midrule
9 & & $\alpha=0$,$~~\beta=0$ & 7 days ahead  & MIN-VAR \\
\addlinespace
10 & Large & $\alpha=0$,$~~\beta=1$ & 7 days ahead  & MIN-VAR \\
\addlinespace
11 & (300 Agents)& $\alpha=0$,$~~\beta=0$ & 7 days ahead  & RMSE \\
\addlinespace
12 & & $\alpha=0$,$~~\beta=1$ & 7 days ahead  & RMSE \\
\bottomrule
\end{tabular}
\label{T:ScenIProf}
\end{table}
%%%%%
%%%%%
%%%%%
Each profile is tested 100 times, and overall there are 1200 experiments in the simulation, and 1200 in the real-world environment. 
To compare the performance between the simulation and live environments, the relative differences between the global cost and average local cost are calculated as:
%%%%%
%%%%%
%%%%%
\begin{equation}
\begin{aligned}
\label{eq:relative}
& \text{\textit{Relative global cost difference:}}~~\frac{g^t_{s,i} - g^t_{l,i}}{g^t_{s,i}} \\
& \text{\textit{Relative average local cost difference:}}~~\frac{l^t_{s,i} - l^t_{l,i}}{l^t_{s,i}} \\
\end{aligned}
\end{equation}
%%%%%
%%%%%
%%%%%
where $g^t_{s,i}$ and $g^t_{l,i}$ are the global costs of profile $i$ at iteration $t$ in simulation and live settings, respectively.
Similarly, $l^t_{s,i}$ and $l^t_{l,i}$ are the average local costs of all agent in profile $i$ at iteration $t$ in simulation and live settings, respectively.
%%%%%
%%%%%
%%%%%
\subsubsection{DIAS} 
\label{S:MethScenIDIAS}
%%%%%
%%%%%
%%%%%
These experiments are based on the GDELT (\textit{Global Dataset of Events, Languages, and Tone}) platform\footnote{\href{https://www.gdeltproject.org}{https://www.gdeltproject.org} [Last accessed: May 2020]}.
\hlc{GDELT monitors and captures print/broadcast/web-based global news media in real-time.
Its data can be accessed via an API in 15-minute intervals.}
This paper employs the DIAS-GDELT demonstrator~\cite{pournaras2019democratizing}\footnote{\href{http://dias-net.org/dias-gdelt-live/}{http://dias-net.org/dias-gdelt-live/} [Last accessed: May 2020]}.
It fetches GDELT news updates every 15 minutes, extracts the possible states, and sends them to the application agents.
DIAS agents (service agents) are mapped to 28 application agents, each representing a country from GDELT.
Each DIAS agent receives the number of news generated during the last 15 minutes from the application agent, disseminates them in the network, and receives the aggregated total number of news generated by the other agents.
%%%%%
%%%%%
%%%%%
\subsection{Evaluation Scenario II: Handling System Dynamics}
\label{S:MethScenII}
%%%%%
%%%%%
%%%%%
In this evaluation scenario, a set of continuous real-world experiments \hlc{between 24/11 - 24/12/2020} are performed to study the performance of both services under complex and dynamic environments.
Each day is divided into three 8-hour time periods: low, medium, and high intensity, each imposing different rate of change for system dynamics.
%%%%%
%%%%%
%%%%%
\subsubsection{I-EPOS} 
\label{S:MethScenIIEPOS}
%%%%%
%%%%%
%%%%%
The experimental settings for this scenario are shown in Table~\ref{T:IEPOSSettings}.
The I-EPOS service is initialized with 200 agents, and each agent is randomly assigned to one of the 2779 EVs from the EV dataset with 7-days-ahead planning horizon.
During runtime, four dynamics are adopted, each corresponding to a change in system settings:
(i) Agents joining/leaving, (ii) local plan change, (iii) $\alpha, \beta$ (weight) change, and (iv) global cost function change.
The rate of change for each dynamic varies across the intensity periods (Table \ref{T:IntensityRates}), with the high-intensity period incurring the highest number of changes.
%%%%%
%%%%%
%%%%%
\begin{table}[t]
\centering
\caption{Rate of change for dynamics in I-EPOS and DIAS live experiments}
\scriptsize
\begin{tabular}{llll} 
\toprule
\textbf{Service / Parameters} &  \multicolumn{3}{c}{\textbf{Intensity / Rate}} \\
\addlinespace
& \textbf{Low} & \textbf{Medium} & \textbf{High} \\
\midrule
\textbf{I-EPOS} & \\
\midrule
Plan Change& $10\%$ & $20\%$ & $50\%$ \\
\addlinespace
$\alpha$ and $\beta$ Change & $10\%$ & $20\%$ & $50\%$ \\
\addlinespace
Global Cost Function (System-wide) & $10\%$ & $20\%$ & $50\%$ \\
\addlinespace
Agent Join/Leave & $10\%$ & $20\%$ & $50\%$ \\
\midrule
\textbf{DIAS} & \\
\midrule 
Change of Possible States& 3h & 2h & 1h \\
\addlinespace
Change of Selected State & $5'$ & $2'$ & $1'$ \\
\addlinespace
Agent Join/Leave & $10'$ & $5'$ & $2'$ \\
\addlinespace
\bottomrule
\end{tabular}
\label{T:IntensityRates}
\end{table} 
%%%%%
%%%%%
%%%%%
For example, in the low-intensity period at the end of each run\footnote{Each run refers to the completion of 50 learning iterations by I-EPOS.} agents change their plans with $5\%$ probability.
The rate of change for $\alpha$ and $\beta$ operates the same way, however, the change in the global cost function is applied system-wide.
The effect of such dynamic changes on the performance of I-EPOS is studied using two metrics:
(i) \hlc{The latency indicates the variation of the I-EPOS execution time with varying dynamics, with respect to non-changing dynamics}~\cite{kaddoum2010criteria}.
The execution time is defined as the time it take (in milliseconds) for I-EPOS to complete 50 iterations, plus applying changes enforced by the dynamics (e.g., agents join/leave, changes in plans, or $\alpha, \beta$ values).
The latency is calculated as follows:
%%%%%
%%%%%
%%%%%
\begin{equation}
\label{eq:latency}
	Latency := \frac{\text{\textit{Varying Dynamics Execution Time}}}{\text{\textit{Non-changing Dynamics Execution Time}}}
\end{equation}
%%%%%
%%%%%
%%%%%
(ii) WAT, which indicates if the system spends excessive time adapting to dynamic changes with respect to performing service-related task.
The WAT is calculated as follows:
%%%%%
%%%%%
\begin{equation}
\label{eq:WAT}
WAT := \frac{\text{\textit{Working~time}}}{\text{\textit{Adaptivity~time}}}
\end{equation}
%%%%%
%%%%%
%%%%%
where the working time concerns the time (in milliseconds) required to execute the 50 learning iterations of I-EPOS, while the adaptivity time concerns the time required to adapt to dynamic changes~\cite{kaddoum2010criteria}.
For instance, adapting to changes in the number of agents, which triggers the self-reorganization of the tree-topology.
%%%%%
%%%%%
%%%%%
\begin{figure*}[t]
\centering
\includegraphics[width = \textwidth]{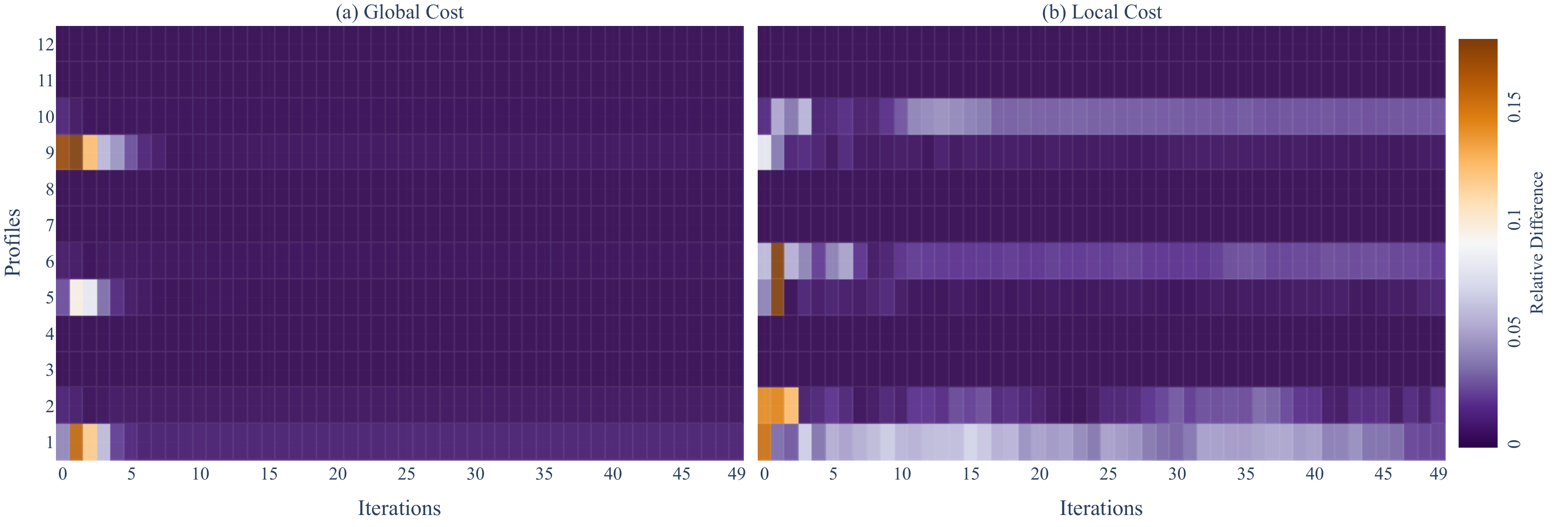}
\caption{Relative difference in global and local cost between the simulation and live settings, calculated based on Equation~\ref{eq:relative}.
The value of each cell shows the mean across 100 repeated experiments for the given profile.
The low values confirm that utilizing the conceptual architecture and the \hlc{Livepeer} toolkit, I-EPOS can transition from simulation to live with minimal introduced error.}
\label{fig:heatmaps}
\end{figure*}
%%%%%
%%%%%
%%%%%
\begin{figure*}[!htb]
\centering
\includegraphics[width = 0.95\textwidth]{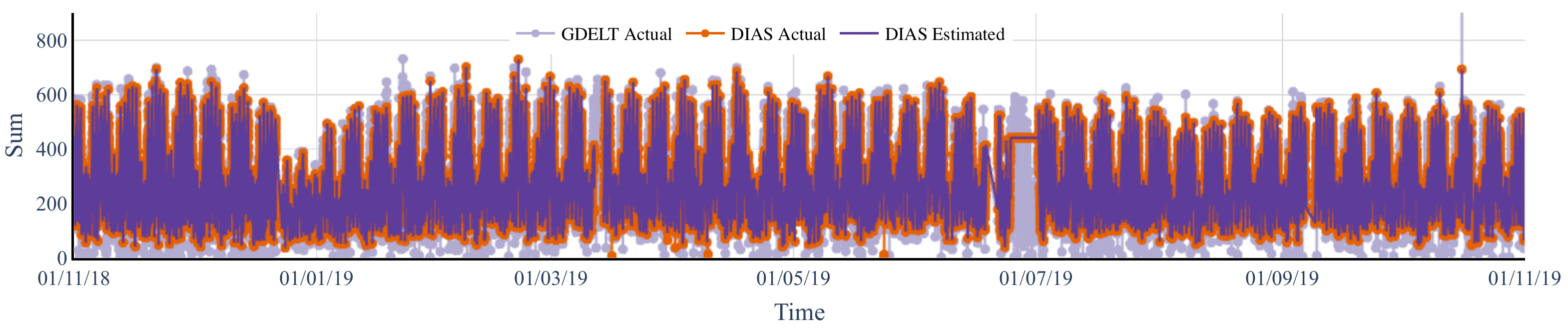}
\caption{(i) GDELT Actual: baseline values extracted from GDELT (i.e., total number of news items generated by the 28 countries)
(ii) DIAS Actual: sum of selected states from each of the 28 DIAS agents, based on the set of possible states for each agent.
(iii) DIAS Estimated: the estimated total number of news items by all countries, calculated by averaging the estimation of each agent.
DIAS can accurately estimate the actual GDELT events in the long term.}
\label{fig:DIAS-GDELT}
\end{figure*}
%%%%%
%%%%%
%%%%%
\subsubsection{DIAS}
\label{S:MethScenIIDIAS}
%%%%%
%%%%%
%%%%%
At the start of each day, DIAS is initialized with 20 agents.
During runtime, three different dynamics are adopted, each corresponding to a change in the system settings:
(i) Agents joining/leaving, (ii) change in the set of possible states, and (iii) change in the selected state.
Every time an agent changes its set of possible states, it randomly selects 9 numbers between the current time and the next hour.
For instance, the possible states for an agent at 14:00 (1400) is 9 numbers in the range of $[1400,1500]$.
The rate of changes for each dynamic varies across the intensity periods (Table \ref{T:IntensityRates}), with the high-intensity period incurring the highest number of changes on the system.
For example, in the low-intensity period, each DIAS agent changes its selected state every 5 minutes.
The agent join/leave rate means that, in the high-intensity case, all agents leave the network every 2 minutes, and return 2 minutes later.
%%%%%
%%%%%
%%%%%
\begin{figure*}[!htb]
\centering
\subfloat{\includegraphics[width = 0.83\textwidth]{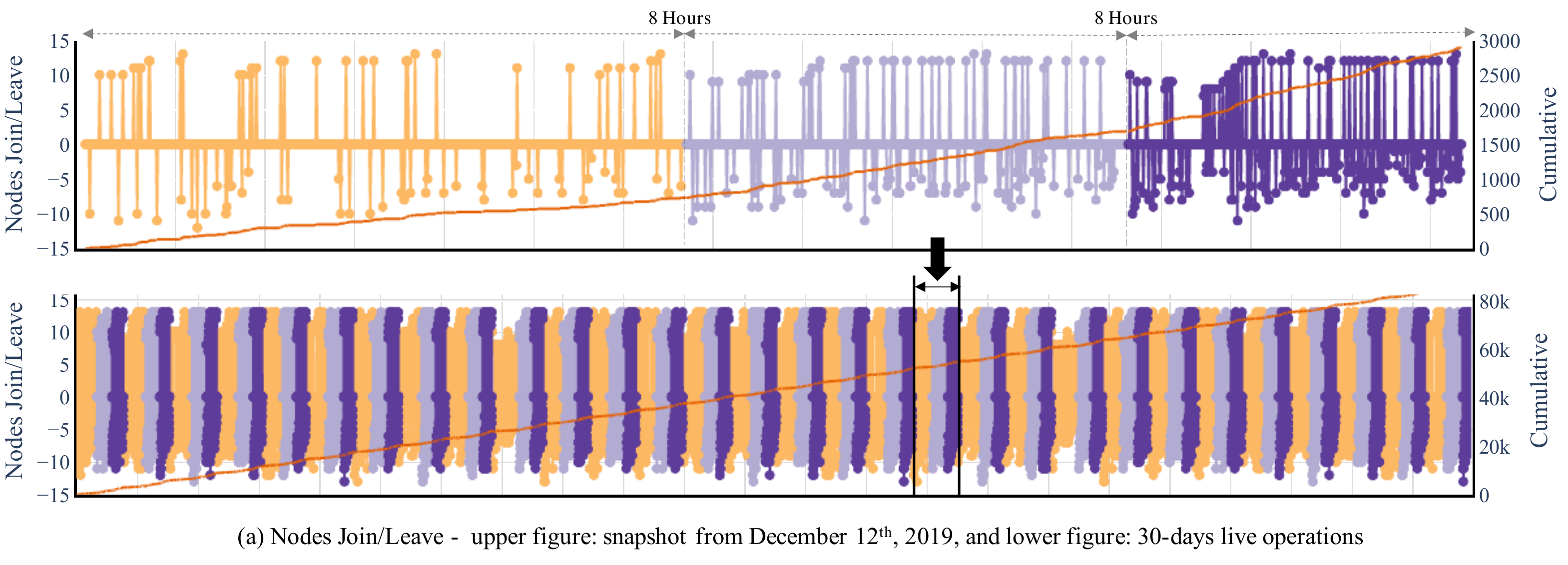}} \\[-0.2ex]
\subfloat{\includegraphics[width = 0.83\textwidth]{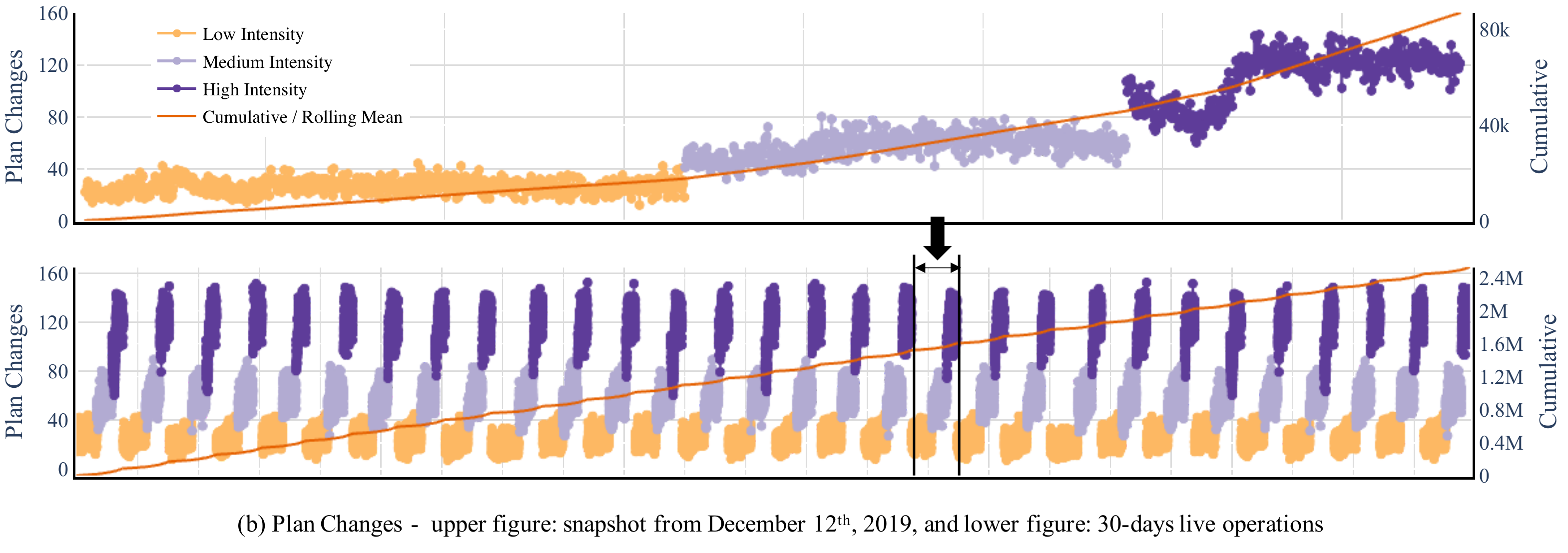}} \\[-0.2ex]
\subfloat{\includegraphics[width = 0.83\textwidth]{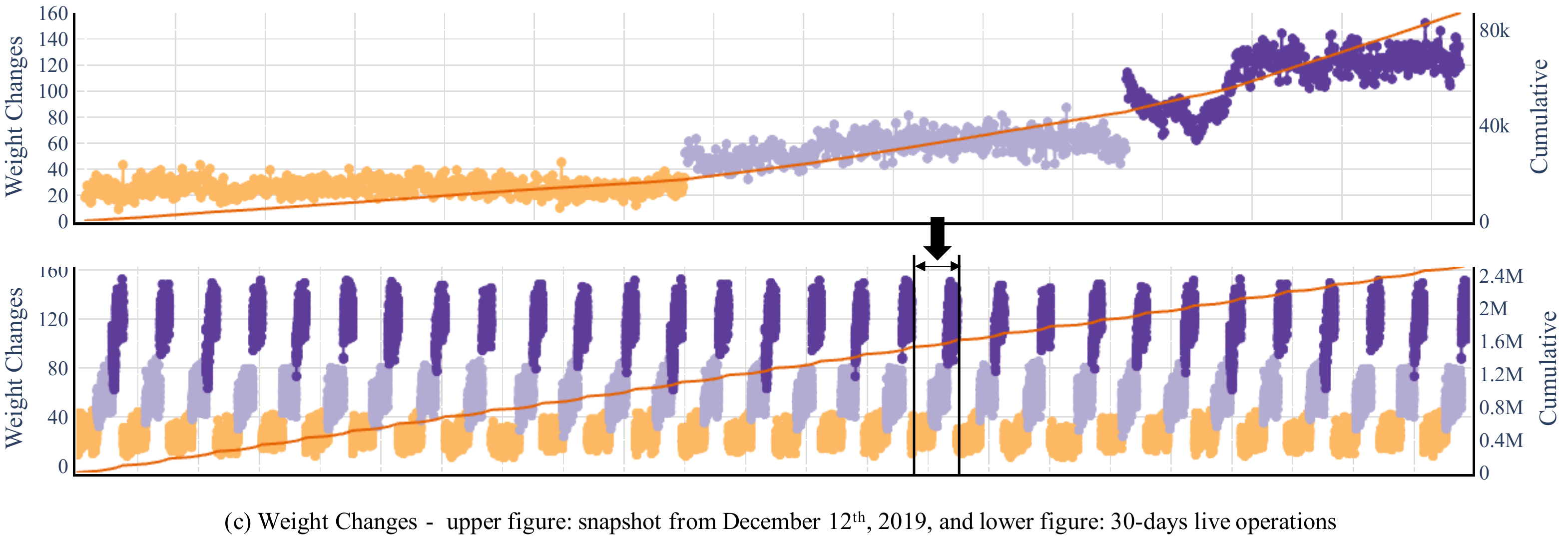}} \\[-0.2ex]
\subfloat{\includegraphics[width = 0.83\textwidth]{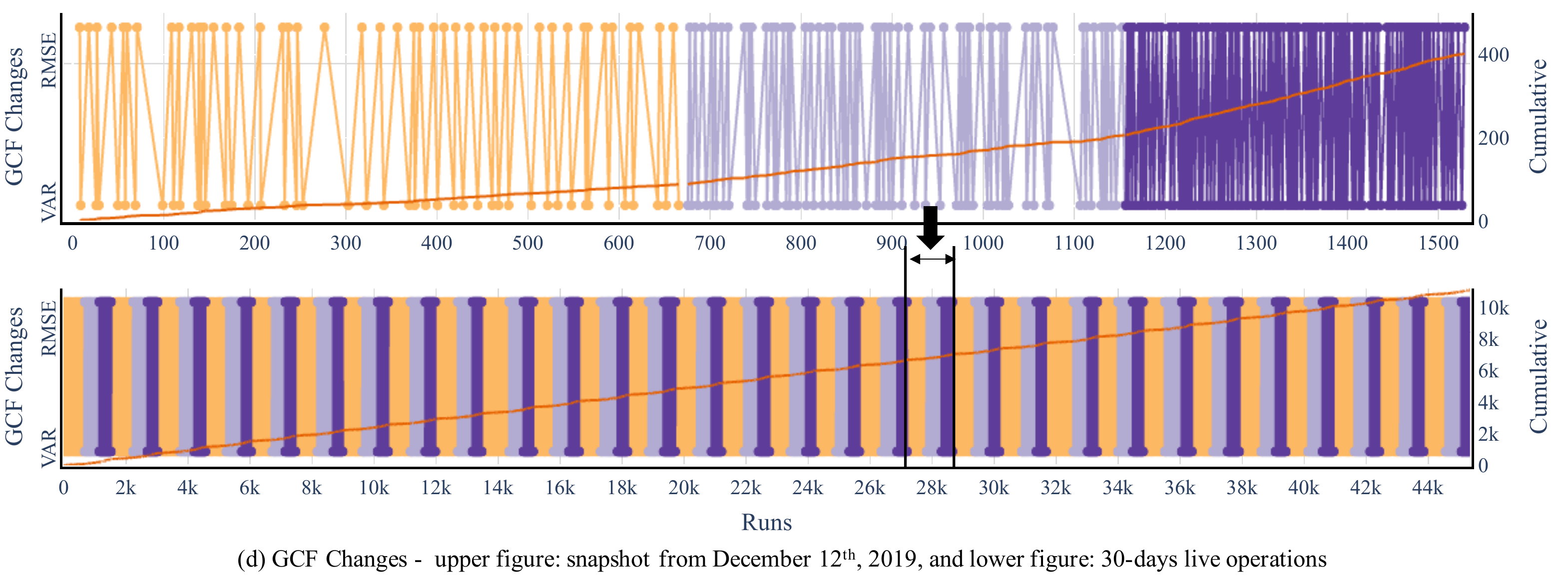}} 
\caption{\hlc{EPOS live operations between 24/11 - 24/12/2020.}
Each run refers to completion of 50 learning iterations by I-EPOS, and GCF denotes the changes in the global cost function, as a system-wide parameter.
\hlc{For each pair, the upper figure shows the snapshot of operations on December $12^{th}$, 2019, while the lower figure illustrates the operations over the month-long experiments.}}
\label{fig:liveEPOS-1}
\end{figure*}
%%%%%
%%%%%
%%%%%
\begin{figure*}[!htb]
\centering
\subfloat{\includegraphics[width = 0.83\textwidth]{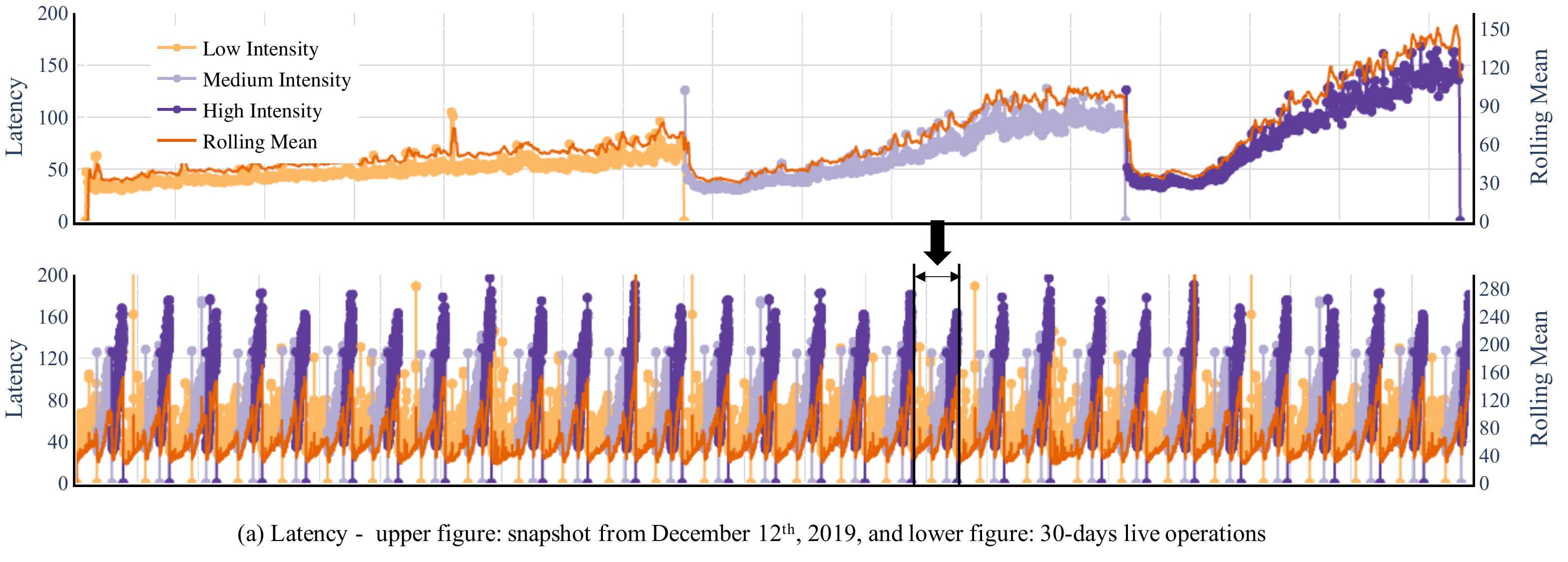}} \\[-0.2ex]
\subfloat{\includegraphics[width = 0.83\textwidth]{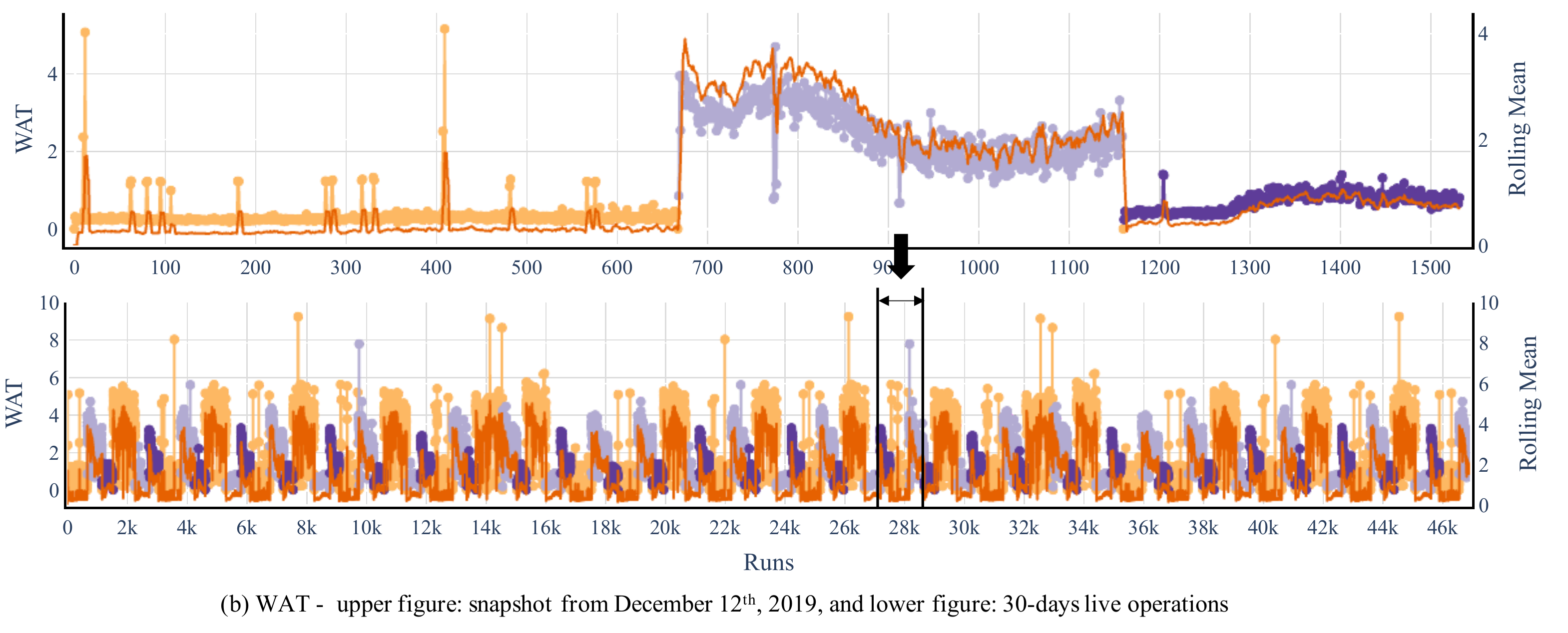}}
\caption{\hlc{EPOS live operations between 24/11 - 24/12/2020, with snapshot on December $12^{th}$, 2019.
Latency indicates the ratio between I-EPOS runtime given varying dynamics with respect to runtime with static non-changing dynamics} (Equation~\ref{eq:latency}).
WAT indicates if the system is spending too much time adapting to dynamic changes rather than performing the I-EPOS learning iterations (Equation~\ref{eq:WAT}).
Note that due to lower WAT (higher adaptivity time), I-EPOS manages to complete fewer runs in higher intensity period, during the same time-frame.}
\label{fig:liveEPOS-2}
\end{figure*}
%%%%%
%%%%%
%%%%%
\begin{figure*}[!htb]
\centering
\subfloat{\includegraphics[width = 0.83\textwidth]{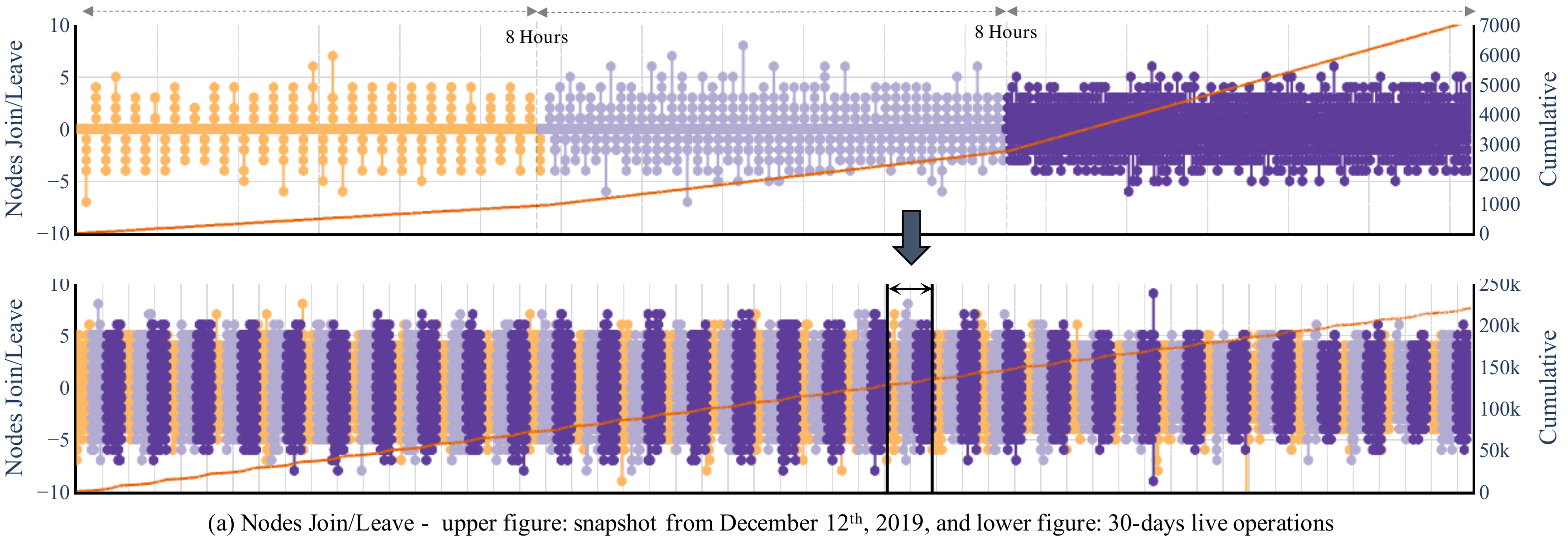}} \\[-0.2ex]
\subfloat{\includegraphics[width = 0.83\textwidth]{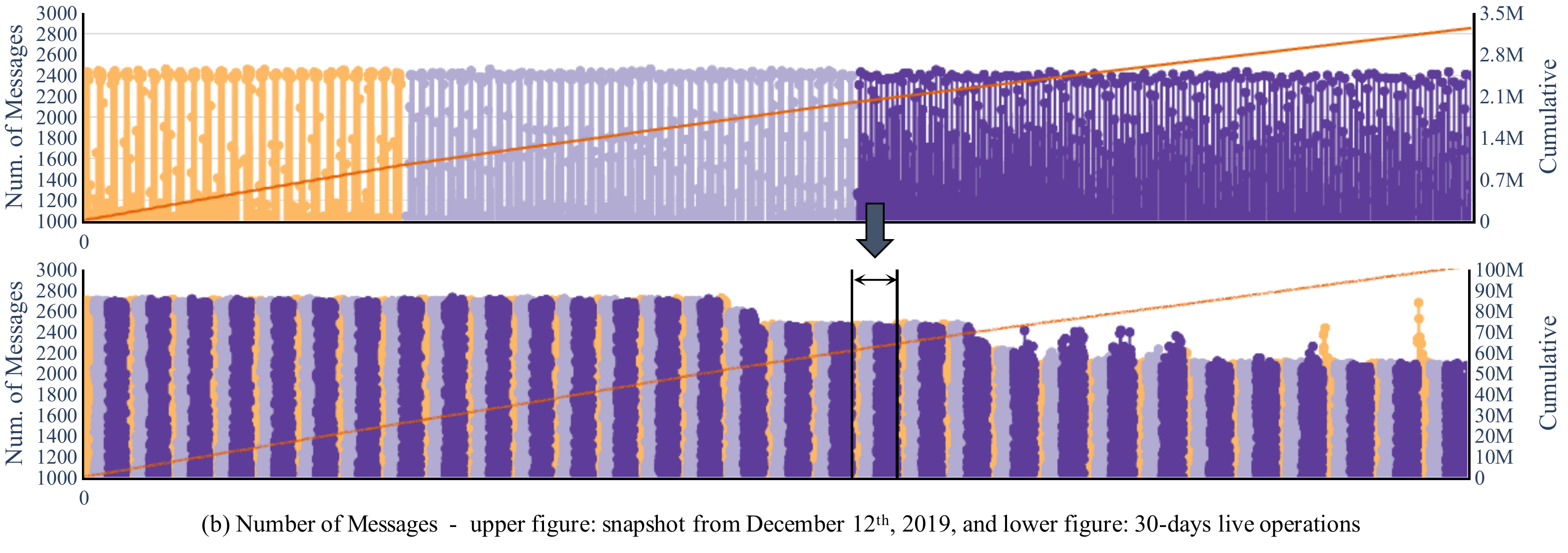}} \\[-0.2ex]
\subfloat{\includegraphics[width = 0.83\textwidth]{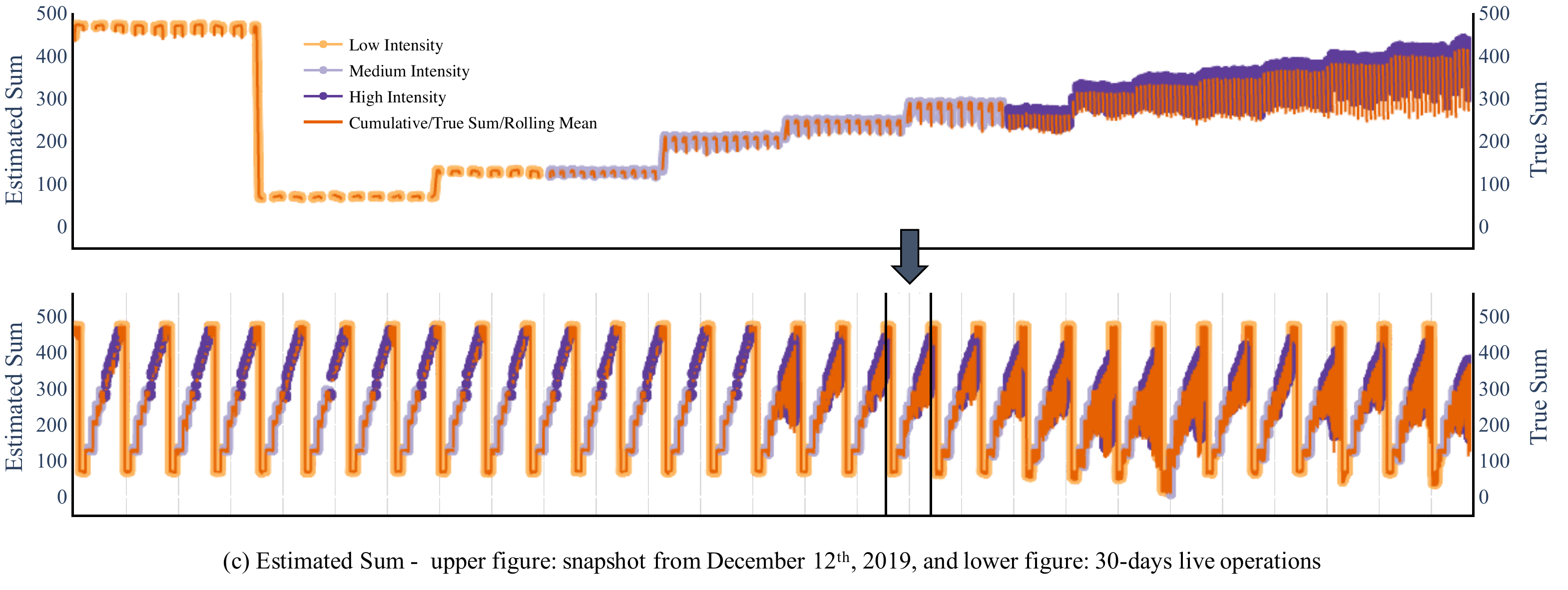}} \\[-0.2ex]
\subfloat{\includegraphics[width = 0.83\textwidth]{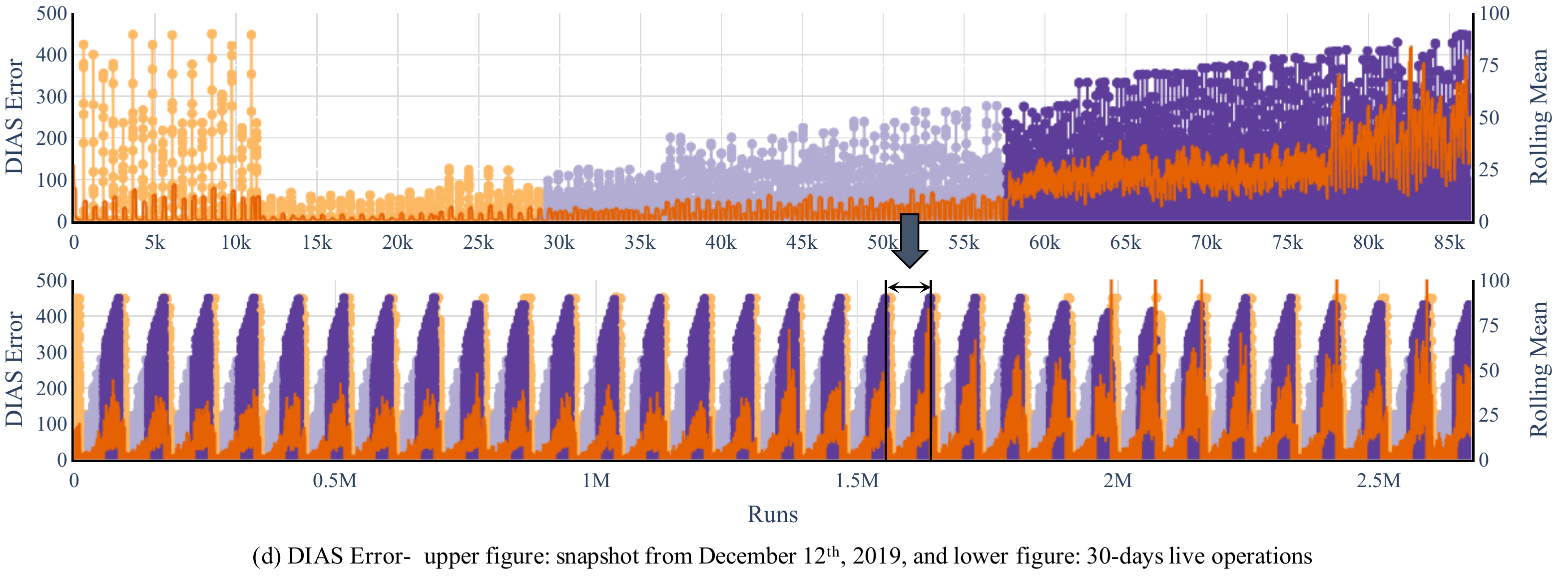}} 
\caption{\hlc{DIAS live operations between 24/11 - 24/12/2020.
For each pair, the upper figure shows the snapshot of operations on December $12^{th}$, 2019, while the lower figure illustrates the operations over the month-long experiments.}
The initial burst in DIAS error (Figure~\ref{fig:liveDIAS}c) is due to the change in set of possible states at midnight.
The estimated sum is calculated by averaging the estimation of each agent, indicating the value each DIAS agent estimates as the true sum of selected state for all DIAS agents.
DIAS error is the difference between true sum (raw data) and the estimated sum.
While this error increases with the rise in intensity, due to quick dissemination and convergence, the rolling mean remains low.}
\label{fig:liveDIAS}
\end{figure*}
%%%%%
%%%%%
%%%%%
\subsection{Deployment Infrastructure}
\label{S:liveDeployment}
%%%%%
%%%%%
%%%%%
The deployment infrastructure of the testbed is as follows:
one server with higher computational power for scaling, and a less powerful server for long-running experiments. 
Both servers provide `bare metal' access, which is substantially faster than using virtual images.
The larger machine has the following specifications:
Intel Xeon hex-core 3.50GHz 256GB DDR3 RAM, 2TB Raid 1 storage, Ubuntu 16.04.
As for the smaller machine: 
Intel Core i7-6700 Quad-Core, 64 GB DDR4 RAM, 1TB storage space, Ubuntu 16.04.
Each service agent is implemented using the \hlc{Livepeer} toolkit (Section \ref{S:DISC}) as a separate JVM object\footnote{By using this approach, in principal agents can run on a different machines and networks.}.
The logging gateway is a single persistence daemon, that creates a single connection to the database (PostgreSQL 10.6) and has a predefined commit rate and queue size that can be adapted based on system scale.
It listens to ZeroMQ messages with logging information sent by the agents, and commits the information to the database.
Each agent notifies the logging gateway of the required relations, tables, and indices, in the form of SQL query templates created on the database.
The communication between all agents is based on message passing, implemented based on the ZeroMQ library.
%%%%%
%%%%%
%%%%%
\section{Experimental Results}
\label{S:ExpEval}
%%%%%
%%%%%
%%%%%
\subsection{Evaluation Scenario I: Comparing Accuracy in Non-volatile Environments}
\label{S:EvalScenI}
%%%%%
%%%%%
%%%%%
This section illustrates the results of the experiments based on the methodology introduces in Section \ref{S:MethScenI}.
%%%%%
%%%%%
%%%%%
\subsubsection{I-EPOS}
\label{S:EvalScenIEPOS}
%%%%%
%%%%%
%%%%%
Figure~\ref{fig:heatmaps} shows the relative difference in global and local cost between the simulation and live environments across the I-EPOS iterations.
For each profile, the global cost is calculated based on the corresponding global cost function (MIN-VAR/MIN-RMSE) in Table~\ref{T:ScenIProf}, and the local cost is the average discomfort of agents' selected plans, calculated by the likelihood of using the EV while it is charging~\cite{pournaras2019socio}.
The value of each cell in Figure~\ref{fig:heatmaps} shows the mean across 100 repeated experiments for the given profile.
\hlc{For each experiment, I-EPOS initializes random trees and assigns the agents to the nodes.
This random assignment generates small variations in the I-EPOS outcome}~\cite{nikolic2019structural}.
However, the general trend across all profiles shows that as the number of learning iterations progresses to the final iteration (50), the global and average local costs of the simulation and live environments converge.
After 10 learning iterations, the relative difference in global cost for all profiles is less than 0.02, and the highest difference in average local cost at the final iteration is 0.0256 related to Profile 10.
This confirms that the I-EPOS transition from simulation to live based on the \hlc{Livepeer} toolkit, can be performed with minimal introduced inaccuracies.
%%%%%
%%%%%
%%%%%
\subsubsection{DIAS}
\label{S:EvalScenIDIAS}
%%%%%
%%%%%
%%%%%
Figure~\ref{fig:DIAS-GDELT} outlines three time-series based on the experimental methodology introduced in Section~\ref{S:MethScenIDIAS}:
(i) GDELT Actual: the raw baseline values extracted from GDELT, representing the total number of news items generated by the 28 countries. 
(ii) DIAS Actual: the sum of selected states from each of the 28 DIAS agents.
Each selected state is the number of generated news items by the assigned country.
The set of possible states is extracted by a sliding a window of 27 observations, uniformly sampling 9 values. 
(iii) DIAS Estimated: The estimated total number of news items by all countries (estimated DIAS actual), calculated by averaging the estimates of each agent.
This estimation is what each DIAS agent calculates as the true value for the DIAS actual.
The accuracy of this estimation is affected by various factors, such as the sampling pool size of data suppliers, convergence time, and the selected state changes.
This experiment has been running since November $1^{st}$, 2018.
As illustrated in Figure~\ref{fig:DIAS-GDELT}, DIAS service based on the \hlc{Livepeer} toolkit can perform long-running operation and accurate estimations of the GDELT baseline. It can rapidly adapt to sudden changes.
%%%%%
%%%%%
%%%%%
\subsection{Evaluation Scenario II: Handling System Dynamics}
\label{S:EvalScenII}
%%%%%
%%%%%
%%%%%
This section illustrates results of the experiments based on the methodology introduced in Section~\ref{S:MethScenII}.
%%%%%
%%%%%
%%%%%
\subsubsection{I-EPOS}
\label{S:EvalScenIIEPOS}
%%%%%
%%%%%
%%%%%
These experiments were continuously executed between 24/11 - 24/12/2019, with the intensity setting changing every 8 hours.
Figures~\ref{fig:liveEPOS-1} and~\ref{fig:liveEPOS-2} illustrate a snapshot of I-EPOS live operation during December $12^{th}$, 2019, \hlc{as well as the live operation during the month-long experiments.}
\hlc{During a typical day (over the month-long period), I-EPOS live handles approximately 80.000 changes in agents' plans (2.4 million), 80.000 changes in $\alpha, \beta$ parameters (2.4 million), 3000 agents joining/leaving (80,000), as well as 400 changes in the global cost function (10,000).}
%%%%%
%%%%%
%%%%%
Figure~\ref{fig:liveEPOS-2}a shows the latency of the I-EPOS across different intensity periods.
On average, the latency increases by $27\%$ from low to medium, and $102\%$ from medium to high.
Figure~\ref{fig:liveEPOS-2}b shows the WAT in different intensity settings, where the average WAT is always higher than 1.
Generally, if the ratio is less than one, the system is spending a lot of time adapting to changes~\cite{kaddoum2010criteria}.
The above experiments confirm that even under highly dynamic environments, I-EPOS completes its learning iterations without any crashes/failures, and delivers the learning outcome.
%%%%%
%%%%%
%%%%%
\subsubsection{DIAS}
\label{S:EvalScenIIDIAS}
%%%%%
%%%%%
%%%%%
The experiments are continuously executed between \hlc{24/11 - 24/12/2019}, with changing intensity settings every 8 hours.
Figures~\ref{fig:liveDIAS}a to~\ref{fig:liveDIAS}d, illustrate a snapshot of DIAS live operation during December $12^{th}$, 2019, \hlc{as well as the live operation during the month-long experiments.}
\hlc{During a typical day (over the month-long period), DIAS handles approximately 4000 agent joins/leaves (250,000), 16,000 state changes (480,000), and 2 million exchanges of messages (100 million).}
The estimated sum of the selected states of all DIAS agents is shown in Figure~\ref{fig:liveDIAS}c.
As shown, even under intense dynamic changes, the DIAS live still provides accurate estimations.
Finally, Figure~\ref{fig:liveDIAS}d illustrates the overall DIAS error, calculated as the difference between true sum (raw data) and the estimated sum.
This error is caused by various factors, such as summarization (raw values to the set of possible states), rapid state changes, agent joining/leaving, and convergence time.
As shown, this error increases with the rise in intensity, however due to quick dissemination of state changes and convergence in the network, the rolling mean error is low.
%%%%%
%%%%%
%%%%%
\section{Conclusion and Future Work}
\label{S:Conclusion}
%%%%%
%%%%%
%%%%%
This paper introduces a novel IoT testbed architecture for decentralized socio-technical services and applications running on IoT.
This architecture applies two layers of abstraction on both the IoT application (devices), and the decentralized services, enabling a dual self-integration capability:
(i) an IoT application integrating several application-independent and modular decentralized services, and
(ii) a decentralized service integrates to several IoT applications without, changing the implementation of the service.
A distributed communication protocol is designed to realize and operationalize the conceptual architecture, providing common interfaces and the communication logic required for self-integration of applications and services at runtime.
Additionally, this paper contributes the \hlc{Livepeer} toolkit, providing a general purpose IoT prototyping toolkit for rapid design and testing of decentralized socio-technical applications, as well as facilitating the transition from simulation to live environments.
Experimental evaluations on two decentralized IoT services, performed under highly dynamic environments confirm the efficiency, and robustness of the testbed architecture. \\

This work promises new instruments for prototyping and developing decentralized socio-technical services running on IoT, and pathways to manage their complexity, which so far have hindered value-oriented self-management approaches in Smart Cities.
Ultimately, the architecture and toolkit will be able to facilitate pilot tests in Smart City use-cases. 
Future research can address the inclusion of other decentralized services with different requirements and network structures, device/agent mobility, and semantic service composition.
Lastly, further deployments in larger-scale infrastructures (e.g., PlanetLab\footnote{\href{https://www.planet-lab.org/about}{https://www.planet-lab.org/about}}) can provide new insights about the applicability of the proposed architecture. 
%while limited access to IoT devices and computational resource narrow down the scale of the experiment to few hundred agents, applying the architecture and toolkit to existing physical testbed, or global research networks (e.g., PlanetLab\footnote{\href{https://www.planet-lab.org/about}{https://www.planet-lab.org/about}}) can provide further insights and improved the applicability of architecture and toolkit.
%%%%%
%%%%%
%%%%%
\section*{Acknowledgement}
This work was supported by the ERC Advanced Grant (324247) Momentum, and the Engineering Social Technologies for a Responsible Digital Future Project at ETH Zurich and TU Delft. Authors would also like to thank Renato Kunz for his contributions in development and implementation of the testbed.
%%%%%
%%%%%
%%%%%
\section*{Artifacts \& Reusability}
\label{S:Reuse}
%%%%%
%%%%%
%%%%%
To facilitate the reusability of the testbed and \hlc{Livepeer} toolkit by the community, the code bases, protocols, and the documentations are made openly available:
Simulation and live versions of I-EPOS\footnote{\href{https://github.com/epournaras/EPOS}{https://github.com/epournaras/EPOS}}, Simulation and live versions of DIAS\footnote{\href{https://github.com/epournaras/DIAS-Development}{https://github.com/epournaras/DIAS-Development}}, monitoring infrastructure\footnote{\href{https://github.com/epournaras/Livelog}{https://github.com/epournaras/Livelog}} and its documentation\footnote{\href{https://github.com/epournaras/Livelog-Documentation}{https://github.com/epournaras/Livelog-Documentation}}, Livepeer\footnote{\href{https://github.com/epournaras/Livepeer}{https://github.com/epournaras/Livepeer}} and its documentation\footnote{\href{https://github.com/epournaras/Livepeer-Documentation}{https://github.com/epournaras/Livepeer-Documentation}}, and IoT device/application agents\footnote{\href{https://github.com/epournaras/DIASClient}{https://github.com/epournaras/DIASClient}} are also available for the community.
%%%%%
%%%%%
%%%%%
\bibliographystyle{model1-num-names}
\bibliography{references}
%%%%%
%%%%%
%%%%%
\end{document}